\documentclass[aps,pra,twocolumn,superscriptaddress,showpacs]{revtex4-1}
\pdfoutput=1\pdfoutput=1
\usepackage[utf8]{inputenc}
\usepackage[english]{babel}
\usepackage[T1]{fontenc}
\usepackage{amsmath}
\usepackage{hyperref}
\usepackage{tikz}
\usepackage{lipsum}
\usepackage{braket}
\usepackage{subfigure}
\usepackage{comment}
\usepackage{mathbbol}
\usepackage{xcolor}
\usepackage{color, soul}
\usepackage{natbib}

\newcommand{\modif}{\color{black}}
\newcommand{\modiff}{\color{black}}

\begin{document}
	
	\title{Atomic interactions for qubit-error compensations}
	
	\author{Michele Delvecchio}
	\email{michele.delvecchio@unipr.it}
	\affiliation{Department of Mathematical, Physical and Computer Sciences, University of Parma, Parco Area delle Scienze 7/A, 43124, Parma, Italy}
	\affiliation{National Institute for Nuclear Physics (INFN), Milano Bicocca Section, Parma Group,  Parco Area delle Scienze 7/A, 43124, Parma, Italy}
	\author{Francesco Petiziol}
	\email{f.petiziol@tu-berlin.de}
	\affiliation{Department of Mathematical, Physical and Computer Sciences, University of Parma, Parco Area delle Scienze 7/A, 43124, Parma, Italy}
	\affiliation{Institut f\"ur Theoretische Physik, Technische Universit\"at Berlin, 
		Hardenbergstr. 36, 10623 Berlin, Germany}  
	\author{Ennio Arimondo}
	\email{ennio.arimondo@unipi.it}
	\affiliation{Dipartimento di Fisica E. Fermi, Universit\`a di Pisa, Largo. B. Pontecorvo 3, 56127 Pisa, Italy}
	\affiliation{INO-CNR, Via G. Moruzzi 1, 56124 Pisa, Italy}
	\author{Sandro Wimberger}
	\email{sandromarcel.wimberger@unipr.it}
	\affiliation{Department of Mathematical, Physical and Computer Sciences, University of Parma, Parco Area delle Scienze 7/A, 43124, Parma, Italy}
	\affiliation{National Institute for Nuclear Physics (INFN), Milano Bicocca Section, Parma Group,  Parco Area delle Scienze 7/A, 43124, Parma, Italy}
	\email{sandromarcel.wimberger@unipr.it}

	\date{\today}
	
	\begin{abstract}
		
		Experimental imperfections induce phase and population errors in quantum systems. 
		We present a method to compensate unitary errors affecting also the population of the qubit states. This is achieved through the interaction of the target qubit with an additional control qubit. We show that our approach works well for single-photon and two-photon excitation schemes. In the first case, we study two reduced models (i) a two-level system in which the interaction corresponds to an effective level shift and (ii) a three-level one describing two qubits in the Bell triplet subspace. In the second case, instead, a double-STIRAP process is presented with comparable compensation efficiency with respect to the single-photon case. 
				
	\end{abstract}
	
	\maketitle 
	
	\section{Introduction}
	\label{sec:introduction}
	
	Quantum computation is based on a series of unitary transformations applied 
	 to computational qubits.  Quantum states are intrinsically delicate with decoherence being the main limitation. In addition, the unitary transformations associated to quantum gates cannot be implemented with perfect accuracy and their small imperfections  will accumulate, leading to a computational failure.  Correction schemes must thus protect against errors. This {\modiff can be done}, for instance, by error correction codes~\cite{NielsenChuang2000, RevModPhys.87.307, PetiziolCarretta2020}, composite pulse sequences~\cite{Levitt1986, Jones2011} or {\modiff {\modif other} robust quantum control protocols} ~\cite{Guery2019, RevQControlEurJPD}. Additional resources are required in all cases: more qubits in the first case, longer times for the qubit preparation and interrogation in the second and additional control fields {\modif or parameter optimization loops} in the latter. Dynamical phase errors can eventually induce also errors of the qubit-level populations. {\modiff Our} ultimate target is to determine and compensate the unwanted phase accumulated in the qubit wavefunction in order to correct coherent (unitary) computational errors. 
	
	\begin{figure}{b}
		\centering
		\includegraphics[width=0.62\columnwidth]{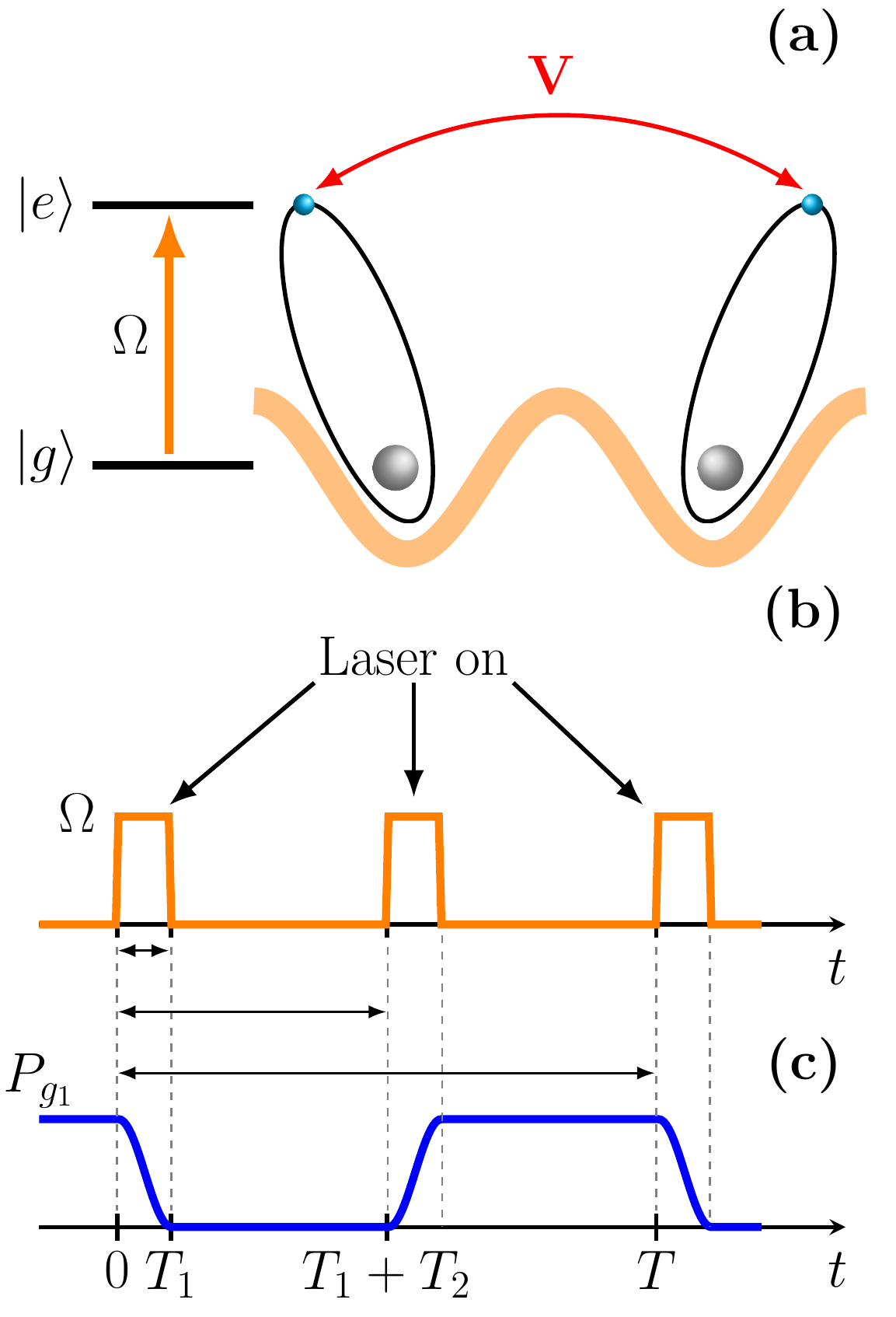}
		\caption{ In (a) scheme of two interacting qubits, for instance, two-level Rydberg atoms trapped in an optical potential and experiencing a nearest-neighbour coupling $V$. They are driven by a common laser field, periodically switched on for the duration $T_1$ and off for the time $T_2$, in a sequence with temporal periodicity $T=2T_1+2T_2$ as schematized in (b). For a perfect excitation, as with $\pi$ pulses for one-photon absorption, the $P_g$ ground state occupation experiences the temporal evolution shown in (c). }
		\label{fig1}
	\end{figure}
		
	The control of the wavefunction phase accumulation has received much attention in different contexts of quantum optics, such as the collapse and subsequent revival of atomic coherence for Bose-Einstein condensates in~\cite{MahmudTsienga2013, MeinertNaegerl2014}. Phase correlation destructions and revivals in the time evolution of dipole-blockaded Rydberg states have been investigated under a detuned and continuous excitation in~\cite{FanWu2019}, and for periodic excitation by~\cite{FanRossini2020} within the context of discrete time crystals. The quantum control of the phase accumulated by laser driving for interacting Rydberg atoms was studied in~\cite{RaoMoelmer2014, BeterovBergamini2020}. An analysis of laser imperfections in the coherent excitation to atomic Rydberg states was experimentally investigated in~\cite{DeLeseleucLahaye2018}.
	
 We introduce here the approach shown in Fig.~\ref{fig1}(a), based on the interaction between the "computation" qubit and an additional "correction" qubit, widely used for the implementation of high fidelity quantum-nondemolition measurements~\cite{WM1995, NielsenChuang2000}.  The idea of using additional qubits for correction is taken from quantum-error-correction protocols and fault-tolerant quantum computing \cite{NielsenChuang2000}, see, {\it e.g.}, \cite{Monroe2021} for a recent experimental realisation using ion qubits. 
	
	In the following, we use the additional phase created by the interaction to compensate for the above unwanted phase in the evolution of the computational qubit. By a proper choice of the interaction strength, we realize compensation for a long sequence of unitary operations of the computational qubit. The compensation efficiency is measured by the wavefunction fidelity reached at the end of the sequence. We obtain a high efficiency for sequence numbers up to fifty.  For the long sequence of applied unitary transformations, we derive conditions for the wavefunction phase to maintain the targeted value.  In most of the  explored qubit level schemes, our approach leads to a magic condition for the required interaction, magic because it creates compensation for a large range of  unitary transformation errors. 
	Our approach is similar to the method of composite pulses {\modiff \cite{Levitt1986,Jones2011}}. In both cases, the phase of the qubit wavefunction accumulated by the laser-pulse sequence produces a more robust qubit excitation. As main difference, the composite-pulse sequence targets a very large and stable fidelity for a single excitation. We target instead a stable and large fidelity in a long sequence of qubit operations.
	Our qubit interaction is linked to the excited state occupation, for instance, in experimental implementations based on atomic Rydberg excitations~\cite{SaffmanMolmer2010, LabuhnBrowaeys2016, Adams2019, Hennrich2020} or on Rydberg-dressed atomic gases~\cite{BorishSchleier-Smith2020}. The interaction can, in principle, be tuned experimentally. Its amplitude depends, {\it e.g.}, on the atomic quantum number, and in addition, in the presence of the F\"orster resonances, it may be controlled by an applied electric field \cite{RavetsBrowaeys2014, Hennrich2017, HuangZheng2018, BeterovBergamini2020}. Similar tunable interactions are present in other realisations as well, {\it e.g.}, in semiconductors \cite{Atac2021} or in artificial atoms, as in the ground state interactions in double quantum dots in a nanowire~\cite{TaherkhaniGregersen2019}. 
	
	\indent  The compensation scheme is applied {\modiff here} to a computational qubit that, under proper laser handling, experiences a Bloch sphere rotation and reinitialization within a given interrogation time interval. We repeat the same sequence on a regular basis, and introduce small errors on the  laser parameters. Therefore, the qubit accumulates an unwanted phase limiting the computation utility. The compensation recovers its utility  through the controlled interaction with the correction qubit. {\modiff {\modif In the following, we specifically consider three situations: two models with a one-photon excitation and one with a two-photon excitation.}
		
	\indent First we consider the one-photon excitation case. Starting from the two qubits sketched in Fig. \ref{fig1}, under the condition for which the laser drive does not influence the correction qubit, the impact of the interaction between the excited levels may be modelled for the computational qubit as an effective shift of the excited level. This gives our system (i): a two-level system exposed to laser pulses driving transitions between the ground and excited state. {\modif In the second configuration, the laser drive couples only to the Bell triplet states and thus the four two-qubit levels can be reduced to three simply by neglecting the Bell singlet state.}
	This three-level system is our model (ii). There is a fundamental difference between these two models: errors in the laser drive can occur only on the computational qubit in (i), while in (ii) both qubits fully participate, with errors that can be modelled on both. }
	 The compensation works {\modiff well} in both cases but the magic value is more stable, producing a larger compensated area with fidelity error of the order of $10^{-6}$, {\modiff for system (i)}. 
	The compensation  is applied to errors in the dynamics of the {\modiff computational} qubit on a Bloch sphere:  rotation angle errors and  rotation axis ones, produced by imperfections in the laser intensity and in the laser detuning, respectively.  We analyze  the one-photon excitation by a $\pi$ laser pulse, allowing a 
	transition at the quantum-speed limit, but with a large sensitivity to the laser excitation parameters~\cite{GiannelliArimondo2014}. The $\pi$ pulses are employed for both the qubit excitation and its reinitialization. 
	
	\indent {\modiff In addition to the one-photon population transfer between ground and upper states, we analyze also a two-photon excitation scheme{ \modif in which the two levels of each qubit cannot directly be coupled by the laser, but are coupled through a third intermediate level. The two-photon excitation analyzed in this case is implemented via a STIRAP in cascade configuration}
	~\cite{VitanovBergmann2017}} with temporal shifted Pump/Stokes laser pulse, providing a large and stable excitation. The application of a double-STIRAP, with inverted Pump/Stokes pulses in the second one, produces the qubit excitation and reinitialization. Similar double-STIRAP pulses are analyzed in~\cite{BeterovSaffman2013} for the implementation of quantum logic gates with Rydberg atomic ensembles.  In several configurations, the three-level STIRAP system may be reduced to a two-level one, where our one-photon compensation scheme may be applied. Instead, we show that the compensation even works in the non-adiabatic regime, where the two level simplification cannot be used~\cite{LaineStenholm1996,VitanovStenholm1996,TorosovVitanov2013}. 
	
	\indent {\modiff The paper is organized as follows: } Sec. \ref{sec:correction} describes the temporal sequence of the computation and reinitialization operations based on the two qubits, the computation and correction one. Such a sequence is repeated for  a longer time period allowing the execution of repeated operations. Sec. \ref{sec:laser_handles} describes the laser driving of the qubits{\modiff{\modif, the corresponding one-photon excitation models} and, finally, the double-STIRAP setup for a two-photon excitation. Moreover, Sec. \ref{sec:laser_handles}} introduces the fidelity as a measure of the compensation efficiency. Section \ref{sec:fid_compensation}  describes the results of the compensation approach, depending on the form of the excitation, based on either single or two-photon processes. {\modiff Sec. \ref{sec:fid_compensation} also presents an alternative and practically useful compensation search tool.  
	 Sec. \ref{sec:conclusion} concludes our work.}

	\section{Qubit correction}
	\label{sec:correction}
	
	We deal with two $j={1,2}$ qubits, the first denoted as "computational" qubit and the second one as "correction" qubit. The computational one performs a quantum computation operation and  repeats its work $n$ times, with $n$ up to 50 in our simulations. Under laser driving, this qubit starts from the initial  ground state $|g_1\rangle$ at time $t=0$, is excited to the state $|e_1\rangle$ to perform its operation, is reinitialized and is ready for the next round at time $t=T$. A not perfect laser driving leads, at time $T$, to computation/reinitialization errors that propagate in the operation sequence. We target to compensate for these  errors.
	
	\indent Our correction process, schematized in Fig.~\ref{fig1}(a), is based  on the  interaction of the first qubit with the correction qubit. These two  qubits, supposed either equal or different, experience  in their  excited states $|e_j\rangle$ an interaction with controllable amplitude $V$ described by the following Hamiltonian in $\hbar$ units:  
	\begin{equation}
		H_V=V|e_1\rangle |e_2\rangle\langle e_1|\langle e_2| \,.
		\label{eq:interaction}
	\end{equation}
	By a proper choice of the interaction amplitude $V$, the phase introduced into the computational qubit  will compensate for laser driving errors.\\
	\indent In our description, the computational process is represented by an elementary step, for instance a rotation of the Bloch vector by a $\pi$ angle~\cite{NielsenChuang2000}. In this simple approach also the reinitialization  is based on a $\pi$ pulse. The computation/correction sequence is based on the following steps, as in Fig. 1(b). At time $t=0$, the computational qubit, initially in its ground state, is transferred to the state $|e_1\rangle$  by a laser pulse of time $T_1$.  In the following time interval $T_2$, the interaction $V$ only determines the qubit evolution. Then an additional  laser pulse of duration $T_1$ transfers the occupation of state $|e_1\rangle$  to the ground state. An interaction of duration $T_2$ completes the sequence with total time $T=2T_1+2T_2$. Under perfect laser driving, the occupation probability of the ground state of the computational qubit follows the time dependence in Fig. 1(c). More precisely, owing to the parity of the ground/excited states, the laser driving takes place through either a one-photon transition or a two-photon one, with a resonant or non-resonant intermediate level. 
	
	
	\indent {\modiff In the following, we will} consider two configurations for the correction qubit in the single-photon excitation case. {\modiff In the first configuration (i), denoted as {\modiff \textit{two-level q-system}} in the following, the control or correction qubit is not exposed to the drive. Under this assumption, the effect of the second static qubit on the computational one is modelled by an effective level shift of its excited state. While in the second configuration (ii), denoted as \textit{three-level q-system}}, the imperfect laser excitation drives both qubits {\modiff simultaneously}, producing an equivalence to the Rydberg blockade. {\modiff The last example we propose is the error compensation in a two interacting three-level setups both driven by the same two-photon STIRAP excitation process.} 
	
	
	
	\section{Qubit laser handles}
	\label{sec:laser_handles}
	\subsection{Single-photon excitation}
	\label{sec:1-phot}
	
	For both the {\modiff two- and three-level} cases, the qubits interact with a laser with detuning $\delta=\omega_L-\omega_0$  from the  $\omega_0$ ground/excited  state transition and Rabi frequency $\Omega$. The $H_{0j}$ Hamiltonian of each atom $(j=1,2)$
	in the rotating wave approximation {\modiff (RWA)} in the frame rotating with the drive, is written in units of $\hbar$ as \begin{equation}
		H_{0j}=- \delta |e_j\rangle\langle e_j|+\frac{\Omega}{2}\left(|g_j\rangle\langle e_j|+|e_j\rangle\langle g_j|\right).
		\label{eq:twolevel} 
	\end{equation}
	{\modiff The RWA, in this case, remains valid in the regime where $\Omega \ll \omega_L $, namely when the operation frequencies we are interested in are much smaller than the carrier frequency of the laser \cite{ShoreBook}. Having in mind typical quantum-optical realizations based on Rydberg atoms this condition is usually satisfied.} \\
	\indent The transfer to the excited states and back is produced by  $\pi$-pulses with $\delta=0$, and $\Omega T_1=\pi$. The ground state occupation reported in Fig.~\ref{fig1}(c) is obtained under these conditions.
	\subsubsection{Two-level q-system (i)}
	\label{sec:one_qubit}
	
	{\modif Here we suppose that the computational and the correction qubit can be addressed independently.}
	{\modiff Then we may assume that} the laser excitation does not influence the correction one,  \textit{e.g.}, due to a large difference in transition frequency of {\modif two different} atoms. For instance, for Rydberg atoms in the presence of F\"orster resonance~\cite{RavetsBrowaeys2014}, we may suppose that the correction qubit remains in a long-lived $|e_2\rangle$ state, the interaction being controlled by switching on/off an electric field. The analysis of the {\modiff two-level q-system (i)} is a very useful step, because it is simpler since the interaction reduces here just to an effective energy shift of the computation qubit. Nevertheless this case leads to the same general compensation scheme valid also for {\modiff our more complex model (ii)}
	
	
	\subsubsection{Three-level  q-system (ii)} 
	\label{sec:two_qubit}
	
	{\modiff As mentioned in the previous section, in this case the laser excitation drives both qubits simultaneously and equally}. Here the Dicke states represent a convenient basis~\cite{FicekTanas2002,ComparatPillet2010,AlmutairiFicek2011} {\modiff for the description}. Starting from the $|g_1,g_2\rangle$ state at $t=0$, only the $|e_1,e_2\rangle$ doubly excited state and the $|s\rangle =(|g_1,e_2\rangle+|e_1,g_2\rangle )/\sqrt{2}$ symmetric states (equivalent to the Bell triplet states) are occupied by the laser excitation with no occupation of the antisymmetric Bell/Dicke state (equivelent to the Bell singlet state). Therefore the wavefunction is written as
	\begin{equation}
		|\psi_{tot}\rangle=c_{gg}|g_1,g_2\rangle+c_{s}|s\rangle+c_{ee}|e_1,e_2\rangle \,,
		\label{eq:Dickestates}
	\end{equation}
	{\modiff and we effectively deal with a three-state system. The Hamiltonian for the two coupled two-level systems (qubits) and its reduction to the symmetric Dicke states is given in app. \ref{appA}.}
	
	
	\indent Let us notice important features playing a key role in the compensation process. Both laser detunings and inter-qubit interaction are diagonal terms{\modiff, see Eq. \eqref{eq:dicke_matrices} in app. \ref{appA}}. From the mathematical point of view, the evolution
	depends only on the sign conformity of  those parameters. This feature will apply also to the compensation results. From a physical point of view the diagonal terms may be used to balance each other. This balance occurs in the Rydberg  blockade~\cite{SaffmanMolmer2010}, where the interaction is large enough to block the laser excitation to the state $|e_1,e_2\rangle$. Viceversa the tuning of those parameters has been used to enhance that excitation as in the ultracold Rydberg  atom antiblockade~\cite{AtosRost2007,AmthorWeidemuller2010} or in the Rydberg enhancement of a room temperature vapor~\cite{KaraMohapatra2018}.  We operate with Rabi frequencies of the laser excitation, much larger than the interaction $V$, where the above processes should \emph{not} occur. 
	
	\subsection{Two-photon excitation}
	\label{sec:two_photon}
\begin{figure}
		\centering
		\includegraphics[width=\columnwidth]{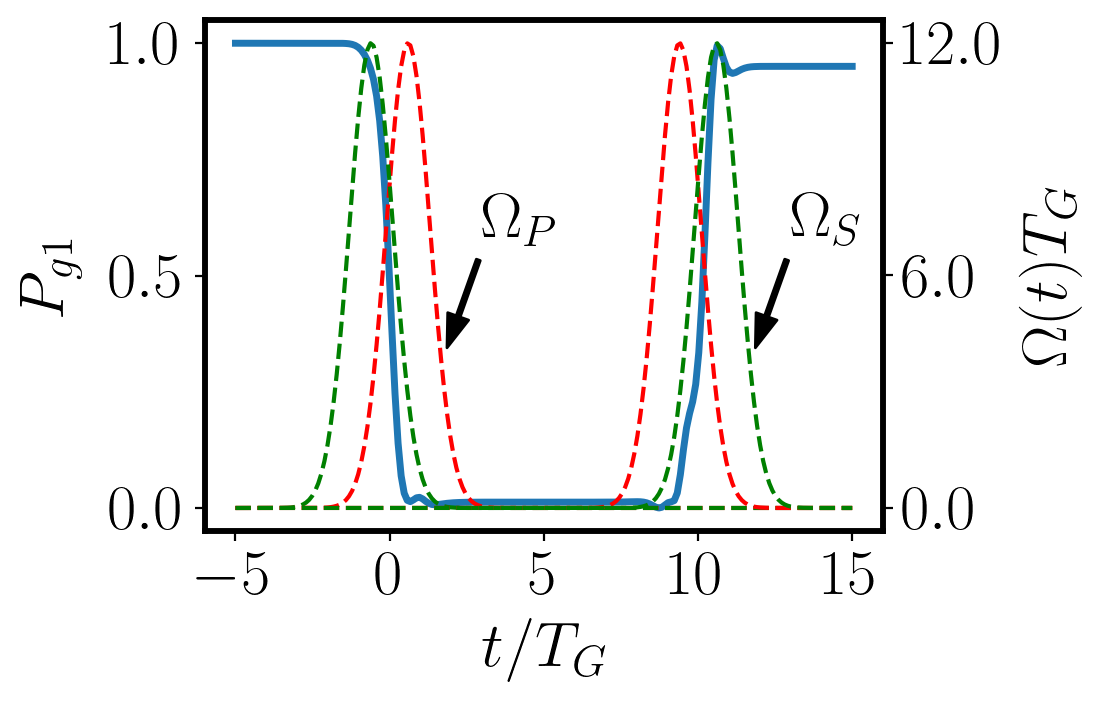}
		\caption{Evolution of the  $P_{g1}$ population (blue solid line), and of the $\Omega_p,\Omega_S$ Rabi frequencies (dashed lines) for a total time $T$, based on a double-STIRAP transfer sequence to the excited state and back to the ground, as for an atomic Rydberg state. Parameters  $T_G=1$, $T_1=1.2$,  $\Delta=\Delta_2=0$, $\Delta T_G=1.4$, $\Omega_0 T_G=12$, and $T_2=10$. The $P_g$ occupation scale is on the left, and the $\Omega(t)T_G$ scale is on the right. The   $P_{g1}$ temporal evolution over a single $T$ sequence  presents the coherent oscillations discussed in~\cite{LaineStenholm1996}. }
		\label{fig:STIRAP}
	\end{figure}

	{\modiff In the previous sections, we have described the single-photon excitation in the two configurations (i) and (ii). In this section, instead, we show the laser handling for a two-photon excitation scheme. In this case,} 
	the $|g_j\rangle$ ground and  $|e_j\rangle$ $(j=1,2)$ excited states{\modiff, that form the qubits, in} a cascade configuration are linked by two laser fields (denoted as Pump and Stokes)  through the intermediate  state $| i_j\rangle$. The Hamiltonian $H_{0j}$ {\modiff representing one of the two qubits is} in a frame doubly rotating at the driving frequencies and in the rotating wave approximation
	\begin{equation} 
		\label{eq:threelevel} 
		H_{j}(t)=\
		\begin{pmatrix}
			0 & \frac{\Omega_P(t)}{2} & 0 \\
			\frac{\Omega_P(t)}{2} & -\Delta & \frac{\Omega_S(t)}{2} \\
			0 &  \frac{\Omega_S(t)}{2} &  -\Delta_2
		\end{pmatrix},
	\end{equation}
	with $\Delta$ the intermediate level detuning, $\Delta_2$ the two-photon detuning, $\Omega_P$ and $\Omega_S$ the Pump and Stokes Rabi frequencies, respectively. For the resonant STIRAP excitation ($\Delta_2=0$) with temporally shifted Pump/Stokes  laser pulses~\cite{VitanovBergmann2017}, the Rabi frequencies have the following Gaussian shaped temporal dependencies:
	\begin{equation}
		\label{eq:gausspulses}
		\Omega_P(t) = \Omega_0 e^{-\left(\frac{t-T_1/2}{T_G}\right)^2}, \quad  \Omega_S(t) =  \Omega_0 e^{-\left(\frac{t+T_1/2}{T_G}\right)^2},
	\end{equation}
	with $\Omega_0$ peak value and width $T_G/\sqrt{2}$. They are parametrized such that the pulse crossing takes place at $t=0$, and the separation between two subsequent maxima is $T_1$.  For the transfer back to the ground state with an inverted Pump/Stokes temporal dependencies,  the pulse crossing occurs at time $t=T_2$. The time sequence of the laser driving is represented with dashed lines in Fig.~\ref{fig:STIRAP}, please see the right vertical scale for the Rabi frequencies. In order to maintain generality with respect to the experimental implementations, we work with STIRAP standard dimensionless quantities $\Omega T_G$ and $\Delta T_G$. This double-STIRAP sequence has been investigated previously, \textit{e.g.}, in~\cite{Hennrich2017, BeterovBergamini2020}.

	\indent An efficient and stable STIRAP transfer is realized by imposing an adiabatic evolution of the dark state $|D_j\rangle$~\cite{Arimondo1996,FleischhauerMarangos2005} defined as
	\begin{equation}
		\label{eq:darkstate}
		|D_j(t)\rangle = \cos \theta(t) |g_j\rangle-\sin \theta(t) |e_j\rangle,
	\end{equation}  %
	with $\theta(t)$ given by $\tan \theta(t) = \Omega_P(t)/\Omega_S(t)$. If the STIRAP parameters  are slowly varied, the qubit initially prepared in the ground state $|g_j\rangle$ follows adiabatically the instantaneous dark state, ending up in the target state $|e_j\rangle$ with very large fidelity. Nonadiabatic coupling between the eigenstates is negligible when the $\theta(t)$ mixing angle rate 
	is smaller than the $\sqrt{\Delta^2+\Omega_P^2+\Omega_S^2}$ separation of the Hamiltonian eigenvalues~\cite{Shore2017}. Robust and efficient STIRAP transfers occur at large values for both Rabi frequencies and intermediate state detuning. There, an adiabatic elimination of the intermediate state leads  to an effective two-level system where our one-photon compensation may be applied. Instead, we investigate a  nonadiabatic regime where the stability with respect to imperfections in the laser parameter is very limited. This occurs for low $\Delta$ values, where the  fidelity presents large oscillations as a function of the laser detuning, see, \textit{e.g.},~\cite{VitanovStenholm1997}. This is seen in Fig.~\ref{fig:STIRAP}, where the fidelity $\mathcal{F}=0.98$ at the end of the first double-STIRAP is not good enough for quantum computation purposes. The complex dependence on $\Delta$ will appear below in Fig.~\ref{fig.STIRAP8compens}(a).
	
	{\modiff While the ground $\ket{g_1}$ and the excited $\ket{e_1}$ states of the three-level STIRAP system are described by Eq. \eqref{eq:threelevel} for the computation qubit, an identical copy forms the control qubit (j=2) that interacts with the first one via an interaction $V$ only between the two excited states, see Eq. \eqref{eq:interaction}. } 

	\subsection{Fidelity/Infidelity}  
	
	The efficiency of the correction qubit approach will be measured at the times $t=nT$ by the fidelity $\mathcal{F}$, or the  infidelity $\mathcal{I}=1-\mathcal{F}$, between the final and the initial (ground) state  $|g_1\rangle$ of the first qubit. For the {\modiff two-state q-system (i), presented in  Sec. \ref{sec:one_qubit},} the straightforward fidelity is, given an arbitrary state of the first qubit $\ket{\psi_1}$,  
	\begin{equation}
		\mathcal{F}_{1}(t=nT)=\left| \langle g_1|\psi_1(nT)\rangle\right|^2.
		\label{eq:fidelity}
	\end{equation}
	
	\indent For the {\modiff three-level setup of Sec. \ref{sec:two_qubit}, instead}, the required fidelity is obtained by performing a reduction (partial trace~\cite{NielsenChuang2000}) to the total 
	density matrix. 
	Using the  $\psi_{\rm tot}$ wavefunction in the Dicke state basis of Eq.~\eqref{eq:Dickestates}, the fidelity  is given by  
	\begin{align}
		\mathcal{F}_{2}&=\text{Tr}[(\ket{g_1} \bra{g_1} \otimes \mathbb{1}_2)(\ket{\psi_{\rm tot}} \bra{\psi_{\rm tot}})] \nonumber \\
		&=|c_{gg} (nT)|^2+\frac{1}{2}|c_s(nT)|^2 \,.
		\label{eq:fid2_dicke}
	\end{align}
	This fidelity contains the predicted occupation of the computational ground state and an additional term associated to the identity of the two qubits. 
	
	\begin{figure}
		\centering
		\includegraphics [width =\columnwidth] {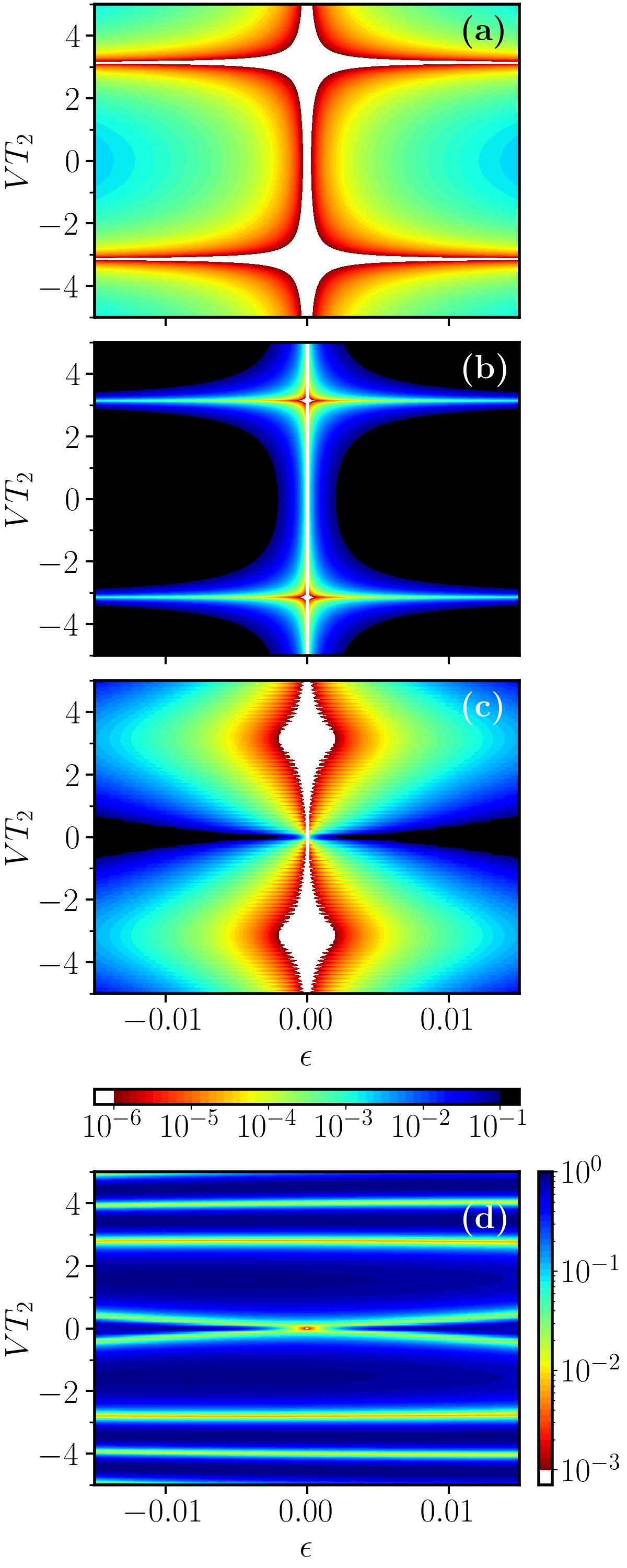}
		\caption{2D contour plots of the logarithmic scale $\mathcal{I}$ infidelity in the ($\epsilon$, $VT_2$)  plane for a rotation error. The colour scale below (c) applies to the plots  (a-c). The black/white regions indicate infidelity values larger/smaller than the color scale limits. Plot of $\mathcal{I}_1(t=T)$ in (a), $\mathcal{I}_{1}(t=50T)$ in (b), and $\mathcal{I}_{2}(t=50T)$ in (c, d).  In (a), and (b) the error compensation corresponds to the configuration {\modiff (i) of Sec. \ref{sec:one_qubit}}, in (c, d) {\modiff the model (ii) of Sec. \ref{sec:two_qubit}}. In (d) the interaction is continuously active.  }
		\label{fig.VErrorRabi}
	\end{figure}
	
	\section{Fidelity compensation}
	\label{sec:fid_compensation}
	In typical applications as considered here, the Rabi frequency $\Omega$ can be much larger than the interaction $V$ {\modiff between control and computational qubit}. In this case, the duration $T_1$ of a $\pi$-pulse is much shorter than the time $T_2$ needed for the interaction to have a sizeable effect. In the limit $T_1 \ll T_2$, it is then justified to treat $V$ as being non-zero only during the interval $T_2$. This assumption enables a simpler analytical treatment adopted in the following expressions, while the general configuration is addressed numerically.
	
	\indent {\modiff We now address typical errors that lead to imperfect population transfer: in Sec.  \ref{sec:rot_ang_err} an error in the rotation angle of the Bloch vector of the form $\sigma_x$, and in Sec. \ref{sec:rot_axis_err} of the rotational axis of the form $\sigma_z$. 
	
	\indent  Given, e.g., the well known long lifetime of the excited Rydberg states \cite{RevModPhys.82.2313}, other effects of decoherence can be safely neglected.}
	
	\subsection{Single-photon excitation}
	
	\subsubsection{Rotation angle error}
		\label{sec:rot_ang_err}
	For the Rabi $\pi$ pulse applied over the $T_1$ time, we introduce a relative error $\epsilon$ given by 
	\begin{equation}
		\Omega T_1=\pi (1+\epsilon)
		\label{eq:Rabierror}
	\end{equation}
	associated to the Bloch vector rotation angle~\cite{NielsenChuang2000}.\\
	\indent  At the time $t=T=2T_1+2T_2$, after the first sequence with two $\pi$ pulses and an acquired phase from the excited state evolution, the ground state fidelity $\mathcal{F}_1(t=T)$ for the {\modiff two-state setup} is
	\begin{eqnarray}
		\mathcal{F}_1(t=T)=& 1-\frac{1}{2}\sin^2\left(2\pi\epsilon\right)\left[1+\cos\left(VT_2\right)\right]  \nonumber \\
		\approx & 1-2(\pi\epsilon)^2\left[1+\cos\left(VT_2\right)\right],
		\label{eq:fidelityRabi}
	\end{eqnarray}
	in the second line at small $\epsilon$ values. The above expressions report an important feature associated to all the compensation schemes here examined. The $VT_2=(2m+1)\pi$ value, with $m$ integer, is "magic" because it produces a strong recovery of the fidelity for the driving error. The excited state  phase shift leads to a positive interference in the wavefunction evolution at all the times $t=nT$. At the magic values, the interaction steps preceding and following the second laser pulse produce the exact reverse of the first pulse rotation angle, as shown in Appendix \ref{appB}. The complete qubit sequence reduces to the identity, and the initial state is exactly restored  for any value of $\epsilon$. While a  full recovery is obtained for all $n$ values at the magic value, Eq.~\eqref{eq:fidelityRabi} reports a fidelity very close to the one for interaction strengths near the magic value. This phase compensation mechanism has a strong analogy with the spin echo and dynamical decoupling techniques \cite{PhysRevLett.82.2417, lidar_brun_2013}, where an arbitrary phase shift is compensated by a double laser excitation.  It is also analog to the anti-resonances of the quantum kicked-rotor model where each pulse exactly counteracts  the preceding one due to a proper choice of free phase evolution in-between the two pulses \cite{Izrailev1990, Sadgrove2011}.  
	
\begin{figure}[t]
		\centering
		\includegraphics [angle=0, width=1 \columnwidth] {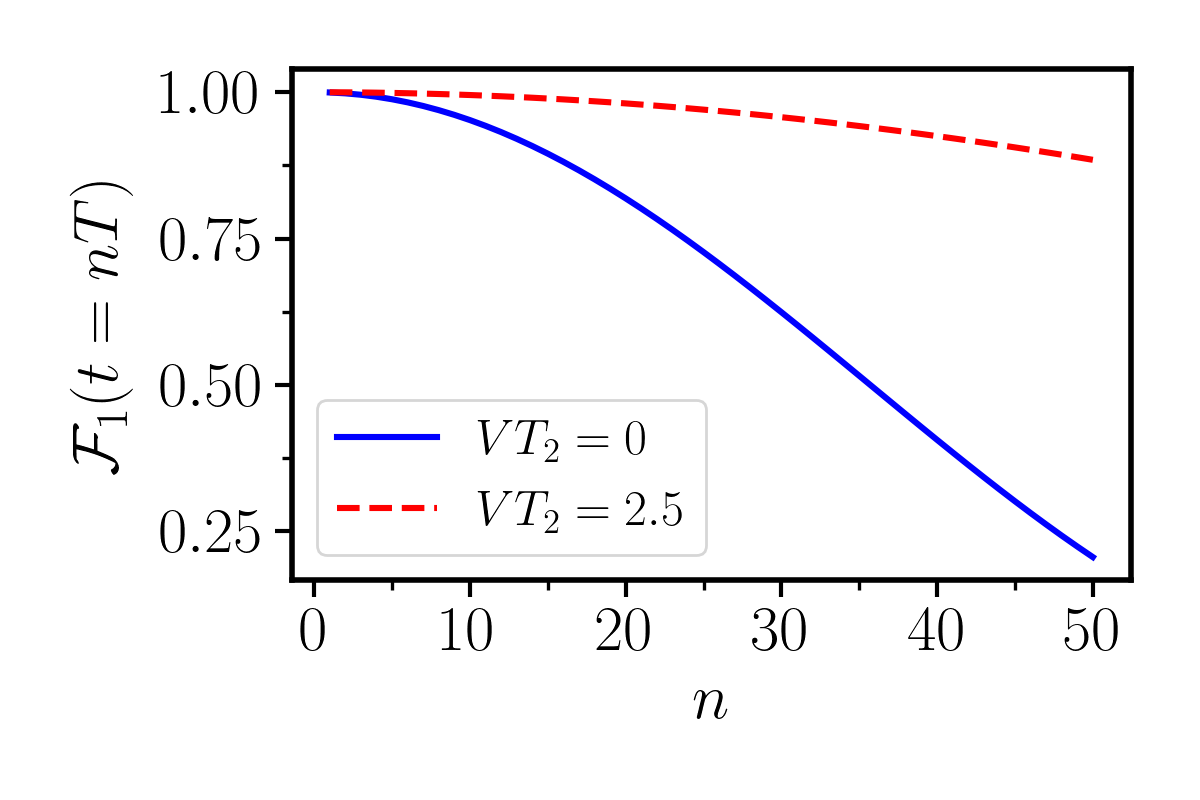}
		\caption{Fidelity $\mathcal{F}_1(t=nT)$ vs. the sequence number $n$ at a given error $\epsilon=0.007$: in the absence of compensation (solid blue line) and for $VT_2=2.5$ (red dashed line) close to the optimal compensation for a laser driving of {\modiff the setup (i)}.}
		\label{fig.FidelityvsN}
	\end{figure}

	\indent With the laser acting on the computational qubit only, {\modiff namely in our model (i),} the plots in the $(\epsilon,VT_2)$ plane report the resulting infidelity  $\mathcal{I}_1(t=T)$ in Fig.~\ref{fig.VErrorRabi}(a) and $\mathcal{I}_{1} (t=50T)$ in (b). These plots, as most of the following ones, are limited on the horizontal to the $|\epsilon | \le0.015$ value, easily reached  in experiments, and to $|VT_2| \le 5$ owing to the vertical periodicity (phase) of the compensation. The infidelity range reported in the plots corresponds to standard quantum computation experiments.  The plots highlight that  the compensation scheme works very well. However, increasing the qubit cycling number, the compensation range around the magic value becomes narrow, and requires a more precise choice of the interaction parameter. The compensation efficiency appears clearly  from the data of Fig.~\ref{fig.FidelityvsN} where the  compensated fidelity $\mathcal{F}_1$ vs. $n$ is compared to the  fidelity reached in multiple operations in the absence of the compensation. For highlighting the robustness of the compensation, notice that the chosen parameter $V$ is only close to the magic value. {\modiff Therefore, a small imperfection in the product $VT_2$ does not affect the efficiency of the compensation.}
	
	\indent {\modiff In the three-level q-model (ii), instead, the} simultaneous driving of both the computation and compensation qubit leads to slightly different features in the time evolution, but the compensation is still produced by the combination of the in-phase laser excitation and the phase acquired here by the state $|e_1,e_2\rangle$ with the laser excitation off. For $V$ off while the laser is on, an analytical solution of the Dicke state time evolution, followed by a projection on the single qubit space, produces the fidelity at time $t=T$ reported in Appendix~\ref{appA} with a complex dependence on sine and cosine functions. At small $\epsilon$ values the fidelity, from Eqs.~\eqref{eq:Cgg} and \eqref{eq:Cs}, becomes
	\begin{equation}
		\mathcal{F}_2(t=T) \approx 1-\frac{1}{2}(\pi\epsilon)^2\left[1+\cos\left(VT_2\right)\right].
		\label{eq:fidelityRabi2Atom}
	\end{equation}
	Only for one rotation at time $t=T$, the phase shift $\pi$ remains magic. 
	However, as the number $n$ of interrogations increases, the behavior of the {\modiff setup (ii)} is different from the case {\modiff (i)}. The fidelity {\modiff of the three-level model} remains overall higher than the one {\modiff of the two-level one}, as can be appreciated from a comparison between the plots (b) and (c) of Fig.~\ref{fig.VErrorRabi}.  
	It should be mentioned here that from the time $t=2T$ onwards the contribution of the Dicke/Bell symmetric state  remains below the $10^{-5}$ level, therefore with an operating  compensation, the computational qubit is in perfect shape for continuing its job.
	
	\indent For the case when the interaction $V$ is applied also within the laser excitation period $T_1$, the full analytical solution is not available. Therefore we rely on numerical simulations, such as shown in Fig.~\ref{fig.VErrorRabi}(d), for the {\modiff three-level model} driving as in all the following figures. The response to the compensation is greatly modified with an optimal compensation that is almost independent of the error $\epsilon$ (for the explored range) at new magic values $VT_2=\pm2.8$ and $VT_2=\pm 3.9$. These values converge to the previous magic ones at larger values of the ratio $T_2/T_1$.
	\begin{figure}[t]
		\centering
		\includegraphics [width=\columnwidth] {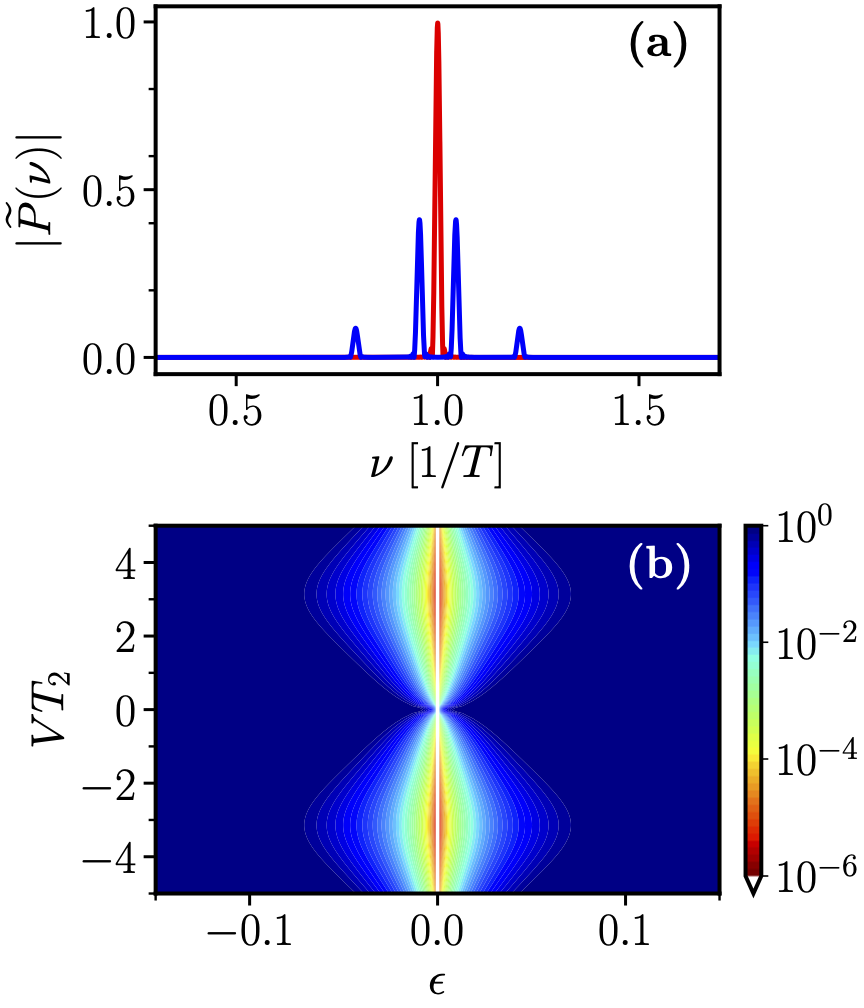}
		\caption{In (a) $|\tilde{P}(\nu)|$ Fourier spectrum of the population difference {\modiff for the two-level setup (i)} at  $\epsilon=0.1$ and for  $VT_2=1$. The sidebands and the splitting are well distinguishable only at these large $\epsilon$ values. They coalesce into $\nu_0$ for the magic compensation value, represented schematically by the red curve.
			In (b) we show $1- \tilde{P}(\nu_0)$ on a logarithmic scale vs. the parameter plane $(\epsilon,VT_2)$ for the rotation error. The white regions indicate infidelity values smaller than the minimum of the color scale.}
		\label{fig.Stroboscopic}
	\end{figure}
	
	\subsubsection{Compensation search tool} 
	Although the study of the fidelity allows us to identify parameter values giving the desired compensation at {\modiff fixed interrogation times} $t=nT$, these values may not guarantee the same fidelity for different $n$. Therefore, we present now a simple approach for determining the compensation parameters guaranteeing high fidelity {\modiff for all periods} $nT$. It is based on the Fourier components of the population difference $P(t) = P_g(t)-P_e(t)$. At $\epsilon=0$ and at perfect compensation,  $P=P_g-P_e$  is a periodic function whose Fourier spectrum contains a single component at $\nu_0=1/T$.  At the $\epsilon \ne 0$ rotation error, a splitting into two sidebands arises in the population difference of the Fourier spectrum $\tilde{P}(\nu)$  at the frequencies $\nu_0(1\pm \epsilon T_1)$, as from the two-level Rabi evolution for $\pi$ pulses. The  interaction $V$ modifies the sideband positions appearing in  Fig.~\ref{fig.Stroboscopic}(a), with a sinusoidal dependence on  $VT_2$. It also introduces additional weaker sidebands at even larger $\epsilon$ values. At the interaction {\modiff values of best} compensation, the main sideband ideally coalesce into the $\nu_0$ value. The $1- \tilde{P}(\nu_0)$ amplitude of the Fourier spectrum peak   vs. the $(\epsilon,VT_2)$ parameters allows a simple determination of the compensation range, as in Fig.~\ref{fig.Stroboscopic}(b). The analogy between this plot and that of Fig.~\ref{fig.VErrorRabi}(c) highligths the utility of this search tool in the full parameter space for the compensation.
	\begin{figure}[t]
		\centering
		\includegraphics [width=0.9\columnwidth] {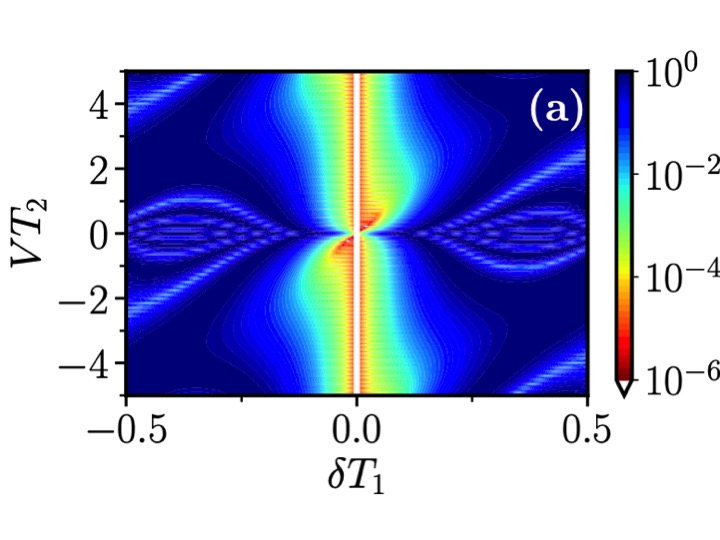}
		\includegraphics [width=0.9\columnwidth] {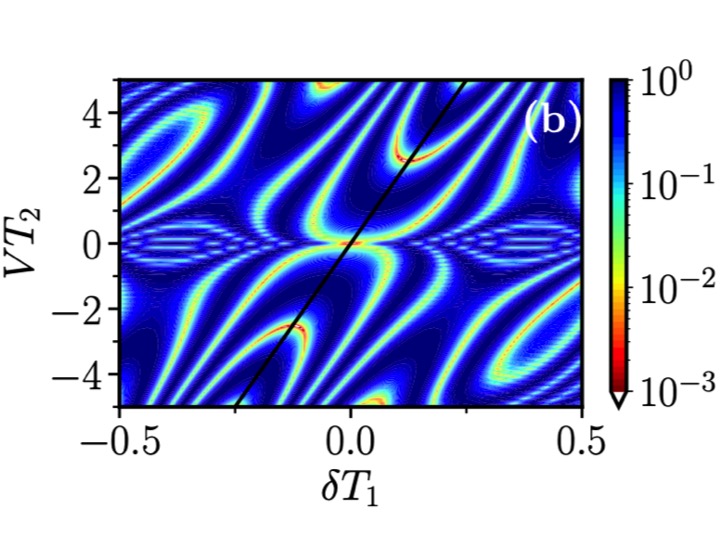}
		\caption{{\modiff Two-level q-system (i): }2D plots of the $\mathcal{I}_{2}(t=50T)$ infidelity plot vs. the $(\delta T_1,VT_2)$ parameters for the laser detuning error. Panel (a) is obtained for the case of $V$ off in the periods when the laser is on. In (b) the interaction is on at all times. Here, the tilted black line corresponds to  the Rydberg enhancement condition $V=2\delta$, see main text.}
		\label{fig.FidelityDelta}
	\end{figure}

	\subsubsection{Rotation axis error} 
	\label{sec:rot_axis_err}
	We examine here the compensation of an error associated to the Bloch vector rotation axis~\cite{NielsenChuang2000}. We suppose that the laser detuning $\delta$ error from resonance is given by 
	\begin{equation}
		\delta T_1=\pi\epsilon \,,
		\label{Deltaerror}
	\end{equation}
 	{\modiff where $\delta T_1 \equiv \delta \cdot T_1$. }
	In a $\pi$ pulse the Bloch vector rotates by the following angle 
	\begin{equation}
		\Omega_{\text{eff}}T_1=\left[\pi^2+(\delta T_1)^2\right]^{1/2} \approx \pi(1+\frac{1}{2}\epsilon^2) \,,
	\end{equation}
	to be compared to Eq.~\eqref{eq:Rabierror} of the previous case. For small $\epsilon$ values, the fidelity at the  time $T$ for the {\modiff model (i)} configuration is
	\begin{equation}
		\begin{split}
			\mathcal{F}_1 (t=T) \approx & 1-2 \epsilon ^2 \left[1-\cos (V T_2 )\right] \\ 
			& -\pi  \epsilon ^3 (2 T_2+1) \sin (V T_2 ) \,.
		\end{split}
		\label{eq:deltafidelity}
	\end{equation}
	Both cosine and sine dependencies on $VT_2$ appear in the present fidelity. These features lead to the compensation  shown in  Fig.~\ref{fig.FidelityDelta}(a). This figure may be compared to Fig.~\ref{fig.VErrorRabi}(c), also for the horizontal scale owing to the detuning error definition of Eq.~\eqref{Deltaerror}. Within the central region of low $\epsilon$ error, the magic $VT_2=\pi$  value does not appear clearly. The  sine term of  Eq.~\eqref{eq:deltafidelity} leads to an asymmetric response of the $(\delta T_1,VT_2)$ plot, breaking the analogy with the Rabi error where a $\delta$ symmetry is valid for all the $\epsilon$ values.
	
	\indent  The results for an  interaction $V$ acting at all times are shown in Fig.~\ref{fig.FidelityDelta}(b). The diagonal line  dependence of the top figure is greatly enhanced.  For a given value of $V$, the compensation is effective within a more limited range of the detuning error. In order to obtain a wider range of error compensation, the role of $V$ within the laser interaction period is reduced by decreasing the ratio $T_1/T_2$.
	
	\indent The diagonal response  in Fig.~\ref{fig.FidelityDelta}(a) for very small values of $\delta T_1 (\approx 0.05$) and in (b) globally, is evidence of the Rydberg antiblockade/enhancement.
	The doubly excited state  for the two atoms is reached for $\delta=V/2$, i.e., $\omega_L=\omega_0+V/2$, taking into account our definition of the interaction energy $V$ provided to the two atoms. This condition corresponds to the black line shown in       
	Fig.~\ref{fig.FidelityDelta}(b). The excitation of both qubits associated to the enhancement allows for the computational qubit to acquire the phase shift  required for the compensation. Notice that, owing to our choice of the laser driving parameters, $\Omega \gg V$, the Rydberg enhancement, and also the blockade, are supposed to be negligible~\cite{SaffmanMolmer2010}. However, even a weak enhancement may contribute essentially to the more sensitive phase-compensation.

	\subsection{Double-STIRAP compensation}
	
	\begin{figure}
		\centering
		\includegraphics [angle=0, width=0.8\columnwidth] {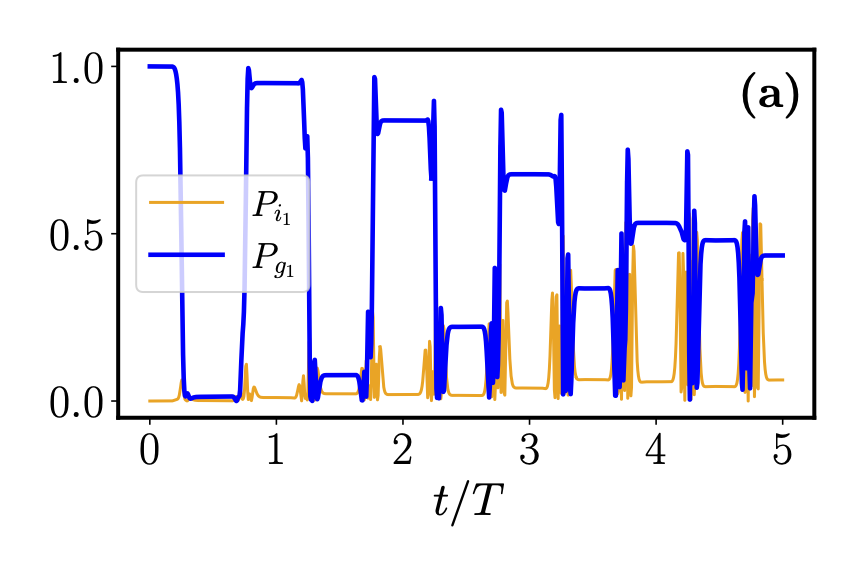}\\
		\includegraphics [angle=0, width=0.8\columnwidth] {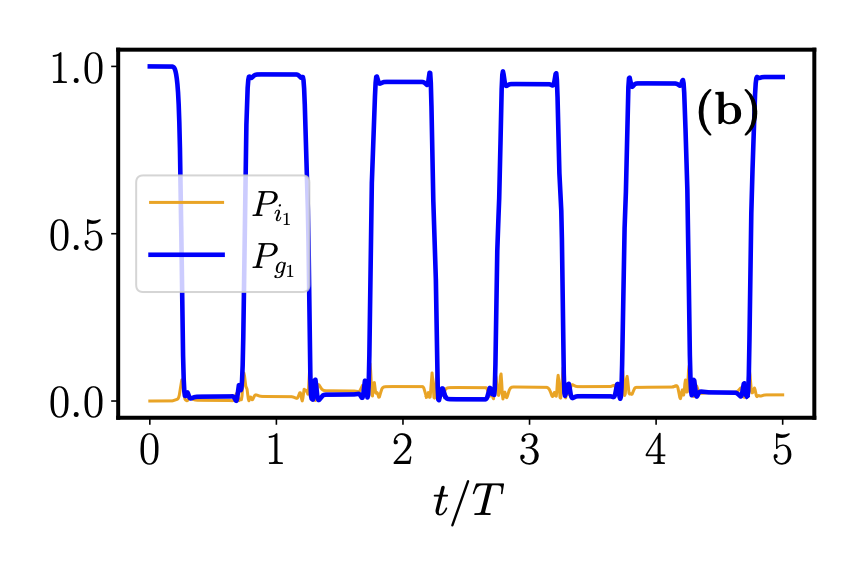} 
		\caption{The  
			$P_{g1}$ population (thick blue solid line) and $P_{i1}$ (thin orange line) vs. $t$ time up to $5T$ for a double-STIRAP operation with $\Delta T_1=1.4$ and parameters as in Fig.~\ref{fig:STIRAP}. (a) without compensation and (b) with an applied compensation $VT_2=2$. Notice the large increase in fidelity produced by the compensation, from $\mathcal{F}_2(t=5T,V=0) \simeq 0.43$ to $\mathcal{F}_2(t=5T,VT_2=2) \simeq 0.96$.}
		\label{fig.STIRAP7compens}
	\end{figure}
\begin{figure}
		\centering           
		\includegraphics [angle=0, width=0.85\columnwidth] {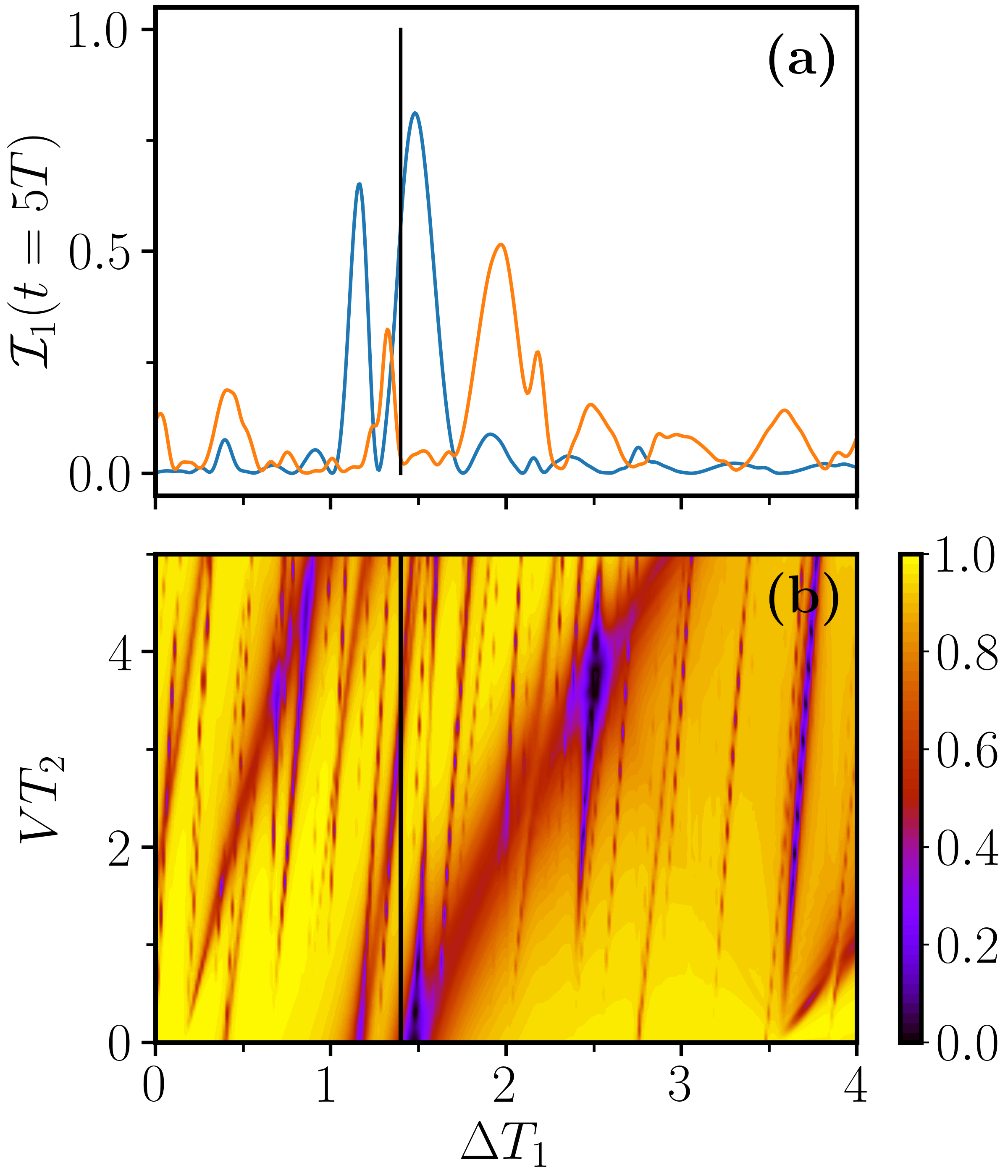}
		\caption{ In (a) $n=5$ infidelities vs. the intermediate level detuning $\Delta$, blue line without compensation and symmetric with respect to the sign of $\Delta$; 
			orange line for an applied compensation $VT_2=2$ and not symmetric with respect to the sign of $\Delta$. The black vertical line marks the $\Delta T_1=1.4$ value of Figs.~\ref{fig:STIRAP} and ~\ref{fig.STIRAP7compens}. The other STIRAP parameters are those given in Fig.~\ref{fig:STIRAP}. (b) two-dimensional $(\Delta T_1, VT_2)$ plot of the $1/T$ Fourier spectrum component for the compensation search. The compensation is efficient in different regions of the parameter plane.  The black vertical line identifies the  $\Delta T_1=1.4$ value, as in (a).} 
		\label{fig.STIRAP8compens}
	\end{figure}

	{\modiff We now turn to the two interacting STIRAP configurations introduced in Sec. \ref{sec:two_photon}.}
	In the case of the two-photon resonant excitation $\Delta_2=0$, we examine the compensation for the non-adiabatic regime in which the fidelity is very sensitive to the  intermediate-level detuning $\Delta$.  As a consequence of the phase accumulation, such a poorly controlled response is enhanced in a long sequence of double-STIRAP pulses. The rapid decrease of the fidelity with the $n$ number, is shown in Fig.~\ref{fig.STIRAP7compens}(a) where the $P_{g1}$ population of the initial state is plotted vs. time up to $t=5T$. Driving  parameters are as in Fig.~\ref{fig:STIRAP}.  An important new feature is the large $P_{i1}$ population occupation of the intermediate state, contributing to less efficient transfer processes with increasing sequence number $n$. The  complex dependence of the infidelity on $\Delta$ is presented in Fig.~\ref{fig.STIRAP8compens}(a) evidencing its drastic increase around the value $\Delta T_1=1.4$. 
	
	\indent We analyze the Fourier spectrum to determine the compensation values when the interaction $V$ is applied at all times. As an example, Fig.~\ref{fig.STIRAP7compens}(b) shows the good and stable fidelity of the computational qubit up $n=5$ when an interaction compensation $VT_2=2$ is applied for the laser parameters of the figure. Very similar results are obtained if the interaction is switched on only for the $T_2$ periods. The two-dimensional plot in Fig.~\ref{fig.STIRAP8compens}(b) for the central Fourier component vs. the $(\Delta T_1,VT_2)$ parameters, shows the presence of regions where the compensation is very efficient. The compensation remains efficient for a reasonable large range of interaction amplitudes. However, for a maximal compensation, a tuning of the  interaction amplitude with the detuning value is required. It should be pointed out that the best compensation is obtained for a sign conformity between $\Delta$ and $V$, as for the data of Fig.~\ref{fig.FidelityDelta}. This response suggests a hidden role of the Rydberg enhancement, in appearance not connected to the $(\delta, V)$ relation discussed above for the one-photon case. In Fig.~\ref{fig.STIRAP8compens}(b)
	the tilted parallel lines corresponding to constant compensation reflect the sinusoidal dependence on the parameter $VT_2$ appearing in the expressions for the fidelity reported above. 
	
	\section{Conclusions}
	\label{sec:conclusion}
	We have introduced a model for robust quantum control, in which {\modiff driving} errors affecting a computational qubit are corrected or compensated via the interaction with an additional correction qubit. For the qubit Bloch vector, the standard errors on rotation angle and axis have been considered. Compensation schemes allowing a very efficient recovery of the fidelity are determined. The fidelity remains very high for up to our choice of fifty qubit operations for the parameters investigated here, but longer sequences could be explored on equal footing. An optimal control approach may be used to determine the compensation in the simultaneous presence of both rotation angle and axis errors.  For ultracold atomic qubits, the interaction is {\modiff naturally} provided by the Rydberg interactions that represent a very efficient tool for quantum simulations or quantum computation \cite{RevModPhys.82.2313, PhysRevLett.85.2208}. Using the F\"orster resonances, the interaction amplitude is easily controlled also with a fast temporal response. 
	{\modif For most of the explored schemes we have demonstrated that the compensation is not very sensitive to the precise value of the interaction strength or, equivalently, the interaction time $T_2$}. In addition the required corrections are within range typically accessible in experimental realizations, and might help to improve substantially the fidelities reached in novel setups such as based on optomechanics \cite{Fedoseev2021}.
	
	\indent The quantum control of a long sequence of qubit operations is determined by numerical calculations. However, the examination of the $t=T$ single sequence by analytical calculation, leads to compensation requirements with a precision good enough for exploring longer temporal sequences. Because the compensation efficiency becomes more critical at larger sequence numbers, it should be tuned performing numerical tests. 
	
	\indent Our scheme has some analogy with the fidelity increase obtained in quantum gates via Rydberg interactions by driving simultaneously the control and target qubits as examined in~\cite{BeterovSaffman2013,SaffmanSanders2020}. It will be interesting to explore the application of our compensation to quantum gates \cite{RaoMoelmer2014, Wu2021, Hennrich2020}. The scheme could also be  extended to more elaborated STIRAP protocols, as that recently introduced in~\cite{DiStefanoFalci2016} for artificial atoms.   An additional feature to be investigated is the role of qubit dissipation within the times of our compensation{\modiff, in particular in view of solid-state, \textit{e.g.}, superconducting qubit implementations \cite{RevModPhys.93.025005}.} 
	
	\section{Acknowledgments}
	One of the authors (E.A.) thanks G. La Rocca and D. Rossini for inspiring discussions. Special thanks goes to G. Falci for a critical reading of the manuscript.
	

	\appendix

\section*{Appendices}

\section{{\modiff Three-level q-system} Hamitonian and fidelity}
\label{appA}

The aim of this Appendix is to derive the Hamiltonian {\modiff for the \textit{three-level q-system}}, once again for the case where the interaction acts only within the $T_2$ periods. The solution in the Dicke basis leads to the fidelity of the computational qubit.

\indent Within the single-qubit basis the Hamiltonian is given by  the sum of $H_{01} \otimes I + I \otimes H_{02}$, with $I$  the identity matrix and $H_{0j}$ from Eq.~\eqref{eq:twolevel}, and of the interaction term of Eq.~\eqref{eq:interaction}. As in Sec.~\ref{sec:correction}, the convenient basis to study the present evolution of two qubits is composed by the three symmetric Dicke states of Eq.~\eqref{eq:Dickestates}. Within that basis the Hamiltonian is described by the following matrices
\begin{align}
	H_0=
	\begin{pmatrix}
		0 & \frac{\Omega}{\sqrt{2}} & 0 \\
		\frac{\Omega }{\sqrt{2}} & \delta   & \frac{\Omega}{\sqrt{2}} \\
		0 & \frac{\Omega}{\sqrt{2}}  & 2 \delta  \\
	\end{pmatrix},  &&
	H_V=
	\begin{pmatrix}
		0 & 0  & 0 \\
		0 & 0   & 0 \\
		0 & 0 & V  \\
	\end{pmatrix} \,,
	\label{eq:dicke_matrices}
\end{align}
for the laser-on periods  and interaction-only periods, respectively. Notice the cooperative $\sqrt{2}$ increase of the Rabi frequency for laser excitation to the  symmetric state $|s\rangle$ with respect to Eqs.~\eqref{eq:twolevel} and \eqref{eq:threelevel}. For $\delta=0$,   the evolution equations are equivalent  to the Bloch equations for a spin-1 system in resonance with the driving field. The interaction produces  a $VT_2$ phase shift of the $|e_1,e_2\rangle$ doubly excited state.

\indent For the calculation of fidelity from Eq.~\eqref{eq:fid2_dicke} the following Dicke state occupations are required:

\begin{align}
	\begin{split}
		|c_{gg}(T)|^2 = &\frac{1}{4}\Big[\left(1-\cos(2\Omega T_1)\right) - \frac{1}{2}\left(1-\cos(\Omega T_1)\right)^2 \times \\ & \times \left(1-\cos(VT_2)\right)\Big]^2+ \\ 
		&+\frac{1}{16} \left[1-\cos(\Omega T_1)\right]^4\sin^2(VT_2)
	\end{split}		 \label{eq:Cgg}\\
	\begin{split}
		|c_s(T)|^2  = & \frac{1}{4}\sin^2(\Omega T_1)\Big[ \cos^2(\Omega T_1) \left(5+3\cos(VT_2)\right) + \\ & + \left(2\cos(\Omega T_1) + 1 \right)\left(1-\cos(VT_2)\right)\Big] \,.
	\end{split} 	\label{eq:Cs}
\end{align}
Using  Eq. \eqref{eq:fid2_dicke} the fidelity at time $(t=T)$ is obtained as 
\begin{align}
	\mathcal{F}_2(t=T)= |c_{gg}(T)|^2 + \frac{1}{2}|c_s(T)|^2 	\,.
\end{align}
Iterating such a procedure, one can obtain analytical expressions of the fidelity at periods $n>1$. However, due to its growing complexity, a software such as Mathematica is helpful for obtaining the rather lengthy algebraic expressions with increasing $n$. {\modiff In the \cite{MathNotebook} we report the Mathematica notebook for the study of the reduced three-level q-system showed in sec \ref{sec:two_qubit}.}

\section{{\modiff Two-level q-system} with $VT_2=\pi$}
\label{appB}

The present target is to derive the magic compensation value from the temporal evolution of the \textit{two-level q-system}, with time separated actions of laser and  interaction. For the case of rotation angle error with an arbitrary Rabi frequency $\Omega$ including the $\epsilon$ error as from Eq.~\eqref{eq:Rabierror}, the evolution operator at time 
$T=2T_1+2T_2$ is 
\begin{equation}
	U(T)=e^{-i VT_2\ket{e_1}\bra{e_1}}e^{-i \frac{\Omega T_1}{2} \sigma_{x1}}e^{-i VT_2\ket{e_1}\bra{e_1}}e^{-i \frac{\Omega T_1}{2} \sigma_{x1}}
	\label{eq:u(T)_1q}
\end{equation}
with $\sigma_{x1}$  the Pauli matrix for the computational qubit. We demonstrate that for $VT_2=\pi$ the operator $U(T)$ corresponds to the identity matrix, for all the $\Omega$ values.

Let us consider the first three exponentials in Eq. \eqref{eq:u(T)_1q} r.h.s.. Written in matrix form, they explicitly read
\begin{equation}
	\begin{split}
		\begin{pmatrix}
			e^{-i VT_2} & 0 \\ 
			0 & 1 
		\end{pmatrix}
		&\begin{pmatrix}
			\cos(\frac{\pi}{2}(1+\epsilon))  && -i\sin(\frac{\pi}{2}(1+\epsilon)) \\
			-i\sin(\frac{\pi}{2}(1+\epsilon))  && \cos(\frac{\pi}{2}(1+\epsilon))
		\end{pmatrix} \times \\
		\times &\begin{pmatrix}
			e^{-i VT_2} & 0 \\ 
			0 & 1 
		\end{pmatrix} \, ,
	\end{split}
\end{equation}
Setting $VT_2=\pi$ and performing the matrix product, we obtain
\begin{equation}
	\begin{pmatrix}
		\cos(\frac{\pi}{2}(1+\epsilon))  && i\sin(\frac{\pi}{2}(1+\epsilon)) \\
		i\sin(\frac{\pi}{2}(1+\epsilon))  && \cos(\frac{\pi}{2}(1+\epsilon))
	\end{pmatrix} 
	= e^{i \frac{\Omega T_1}{2} \sigma_x} \, .
\end{equation}
This operator performs  a rotation of the Bloch vectors equal and opposite to that induced by the first pulse of the sequence in Eq.~\eqref{eq:u(T)_1q}. Inserting the latter result in that equation we obtain
\begin{equation}
	U(T)=e^{i \frac{\Omega T_1}{2} \sigma_x}e^{-i \frac{\Omega T_1}{2} \sigma_x}=I \, ,
\end{equation}
demonstrating the magic value compensation, independent of the $\epsilon$ rotation angle error.

\indent The application of the evolution operator to the case of the rotation axis error unfortunately
does not produce a similar simple interpretation since both errors cannot be exactly corrected by the same procedure from above. 
One may apply, however, an optimal control approach to find the optimal value for the best simultaneous compensation of both errors.


\begin{thebibliography}{55}%
	\makeatletter
	\providecommand \@ifxundefined [1]{%
		\@ifx{#1\undefined}
	}%
	\providecommand \@ifnum [1]{%
		\ifnum #1\expandafter \@firstoftwo
		\else \expandafter \@secondoftwo
		\fi
	}%
	\providecommand \@ifx [1]{%
		\ifx #1\expandafter \@firstoftwo
		\else \expandafter \@secondoftwo
		\fi
	}%
	\providecommand \natexlab [1]{#1}%
	\providecommand \enquote  [1]{``#1''}%
	\providecommand \bibnamefont  [1]{#1}%
	\providecommand \bibfnamefont [1]{#1}%
	\providecommand \citenamefont [1]{#1}%
	\providecommand \href@noop [0]{\@secondoftwo}%
	\providecommand \href [0]{\begingroup \@sanitize@url \@href}%
	\providecommand \@href[1]{\@@startlink{#1}\@@href}%
	\providecommand \@@href[1]{\endgroup#1\@@endlink}%
	\providecommand \@sanitize@url [0]{\catcode `\\12\catcode `\$12\catcode
		`\&12\catcode `\#12\catcode `\^12\catcode `\_12\catcode `\%12\relax}%
	\providecommand \@@startlink[1]{}%
	\providecommand \@@endlink[0]{}%
	\providecommand \url  [0]{\begingroup\@sanitize@url \@url }%
	\providecommand \@url [1]{\endgroup\@href {#1}{\urlprefix }}%
	\providecommand \urlprefix  [0]{URL }%
	\providecommand \Eprint [0]{\href }%
	\providecommand \doibase [0]{https://doi.org/}%
	\providecommand \selectlanguage [0]{\@gobble}%
	\providecommand \bibinfo  [0]{\@secondoftwo}%
	\providecommand \bibfield  [0]{\@secondoftwo}%
	\providecommand \translation [1]{[#1]}%
	\providecommand \BibitemOpen [0]{}%
	\providecommand \bibitemStop [0]{}%
	\providecommand \bibitemNoStop [0]{.\EOS\space}%
	\providecommand \EOS [0]{\spacefactor3000\relax}%
	\providecommand \BibitemShut  [1]{\csname bibitem#1\endcsname}%
	\let\auto@bib@innerbib\@empty
	\bibitem [{\citenamefont {Nielsen}\ and\ \citenamefont
		{Chuang}(2000)}]{NielsenChuang2000}%
	\BibitemOpen
	\bibfield  {author} {\bibinfo {author} {\bibfnamefont {M.~A.}\ \bibnamefont
			{Nielsen}}\ and\ \bibinfo {author} {\bibfnamefont {I.~L.}\ \bibnamefont
			{Chuang}},\ }\href@noop {} {\emph {\bibinfo {title} {{Quantum Computation and
					Quantum Information}}}}\ (\bibinfo  {publisher} {Cambridge University
		Press},\ \bibinfo {year} {2000})\BibitemShut {NoStop}%
	\bibitem [{\citenamefont {Terhal}(2015)}]{RevModPhys.87.307}%
	\BibitemOpen
	\bibfield  {author} {\bibinfo {author} {\bibfnamefont {B.~M.}\ \bibnamefont
			{Terhal}},\ }\bibfield  {title} {\bibinfo {title} {Quantum error correction
			for quantum memories},\ }\href {https://doi.org/10.1103/RevModPhys.87.307}
	{\bibfield  {journal} {\bibinfo  {journal} {Rev. Mod. Phys.}\ }\textbf
		{\bibinfo {volume} {87}},\ \bibinfo {pages} {307} (\bibinfo {year} {2015})},\
	\bibinfo {note} {arXiv:1302.3428v7}\BibitemShut {NoStop}%
	\bibitem [{\citenamefont {Chiesa}\ \emph {et~al.}(2020)\citenamefont {Chiesa},
		\citenamefont {Macaluso}, \citenamefont {Petiziol}, \citenamefont
		{Wimberger}, \citenamefont {Santini},\ and\ \citenamefont
		{Carretta}}]{PetiziolCarretta2020}%
	\BibitemOpen
	\bibfield  {author} {\bibinfo {author} {\bibfnamefont {A.}~\bibnamefont
			{Chiesa}}, \bibinfo {author} {\bibfnamefont {E.}~\bibnamefont {Macaluso}},
		\bibinfo {author} {\bibfnamefont {F.}~\bibnamefont {Petiziol}}, \bibinfo
		{author} {\bibfnamefont {S.}~\bibnamefont {Wimberger}}, \bibinfo {author}
		{\bibfnamefont {P.}~\bibnamefont {Santini}},\ and\ \bibinfo {author}
		{\bibfnamefont {S.}~\bibnamefont {Carretta}},\ }\bibfield  {title} {\bibinfo
		{title} {Molecular nanomagnets as qubits with embedded quantum-error
			correction},\ }\href {https://doi.org/10.1021/acs.jpclett.0c02213} {\bibfield
		{journal} {\bibinfo  {journal} {J. Phys. Chem. Lett.}\ }\textbf {\bibinfo
			{volume} {11}},\ \bibinfo {pages} {8610} (\bibinfo {year} {2020})},\ \bibinfo
	{note} {pMID: 32936660}\BibitemShut {NoStop}%
	\bibitem [{\citenamefont {Levitt}(1986)}]{Levitt1986}%
	\BibitemOpen
	\bibfield  {author} {\bibinfo {author} {\bibfnamefont {M.~H.}\ \bibnamefont
			{Levitt}},\ }\bibfield  {title} {\bibinfo {title} {Composite pulses},\ }\href
	{https://doi.org/https://doi.org/10.1016/0079-6565(86)80005-X} {\bibfield
		{journal} {\bibinfo  {journal} {Progr. Nucl. Magn. Res. Spectrosc.}\ }\textbf
		{\bibinfo {volume} {18}},\ \bibinfo {pages} {61 } (\bibinfo {year}
		{1986})}\BibitemShut {NoStop}%
	\bibitem [{\citenamefont {Jones}(2011)}]{Jones2011}%
	\BibitemOpen
	\bibfield  {author} {\bibinfo {author} {\bibfnamefont {J.~A.}\ \bibnamefont
			{Jones}},\ }\bibfield  {title} {\bibinfo {title} {Quantum computing with
			{NMR}},\ }\href {https://doi.org/https://doi.org/10.1016/j.pnmrs.2010.11.001}
	{\bibfield  {journal} {\bibinfo  {journal} {Progr. Nucl. Magn. Res.
				Spectrosc.}\ }\textbf {\bibinfo {volume} {59}},\ \bibinfo {pages} {91 }
		(\bibinfo {year} {2011})},\ \bibinfo {note} {arXiv:1011.1382v1}\BibitemShut
	{NoStop}%
	\bibitem [{\citenamefont {Gu\'ery-Odelin}\ \emph {et~al.}(2019)\citenamefont
		{Gu\'ery-Odelin}, \citenamefont {Ruschhaupt}, \citenamefont {Kiely},
		\citenamefont {Torrontegui}, \citenamefont {Mart\'{\i}nez-Garaot},\ and\
		\citenamefont {Muga}}]{Guery2019}%
	\BibitemOpen
	\bibfield  {author} {\bibinfo {author} {\bibfnamefont {D.}~\bibnamefont
			{Gu\'ery-Odelin}}, \bibinfo {author} {\bibfnamefont {A.}~\bibnamefont
			{Ruschhaupt}}, \bibinfo {author} {\bibfnamefont {A.}~\bibnamefont {Kiely}},
		\bibinfo {author} {\bibfnamefont {E.}~\bibnamefont {Torrontegui}}, \bibinfo
		{author} {\bibfnamefont {S.}~\bibnamefont {Mart\'{\i}nez-Garaot}},\ and\
		\bibinfo {author} {\bibfnamefont {J.~G.}\ \bibnamefont {Muga}},\ }\bibfield
	{title} {\bibinfo {title} {Shortcuts to adiabaticity: Concepts, methods, and
			applications},\ }\href {https://doi.org/10.1103/RevModPhys.91.045001}
	{\bibfield  {journal} {\bibinfo  {journal} {Rev. Mod. Phys.}\ }\textbf
		{\bibinfo {volume} {91}},\ \bibinfo {pages} {045001} (\bibinfo {year}
		{2019})},\ \bibinfo {note} {arXiv:1904.08448v2}\BibitemShut {NoStop}%
	\bibitem [{\citenamefont {Glaser}\ \emph {et~al.}(2015)\citenamefont {Glaser},
		\citenamefont {Boscain}, \citenamefont {Calarco}, \citenamefont {Koch},
		\citenamefont {K\"ockenberger}, \citenamefont {Kosloff}, \citenamefont
		{Kuprov}, \citenamefont {Luy}, \citenamefont {Schirmer}, \citenamefont
		{Schulte-Herbr\"uggen}, \citenamefont {Sugny},\ and\ \citenamefont
		{Wilhelm}}]{RevQControlEurJPD}%
	\BibitemOpen
	\bibfield  {author} {\bibinfo {author} {\bibfnamefont {S.~J.}\ \bibnamefont
			{Glaser}}, \bibinfo {author} {\bibfnamefont {U.}~\bibnamefont {Boscain}},
		\bibinfo {author} {\bibfnamefont {T.}~\bibnamefont {Calarco}}, \bibinfo
		{author} {\bibfnamefont {C.~P.}\ \bibnamefont {Koch}}, \bibinfo {author}
		{\bibfnamefont {W.}~\bibnamefont {K\"ockenberger}}, \bibinfo {author}
		{\bibfnamefont {R.}~\bibnamefont {Kosloff}}, \bibinfo {author} {\bibfnamefont
			{I.}~\bibnamefont {Kuprov}}, \bibinfo {author} {\bibfnamefont
			{B.}~\bibnamefont {Luy}}, \bibinfo {author} {\bibfnamefont {S.}~\bibnamefont
			{Schirmer}}, \bibinfo {author} {\bibfnamefont {T.}~\bibnamefont
			{Schulte-Herbr\"uggen}}, \bibinfo {author} {\bibfnamefont {D.}~\bibnamefont
			{Sugny}},\ and\ \bibinfo {author} {\bibfnamefont {F.~K.}\ \bibnamefont
			{Wilhelm}},\ }\bibfield  {title} {\bibinfo {title} {Training
			{S}chr\"odinger\'{}s cat: quantum optimal control - strategic report on
			current status, visions and goals for research in europe},\ }\href
	{https://doi.org/10.1140/epjd/e2015-60464-1} {\bibfield  {journal} {\bibinfo
			{journal} {Eur. Phys. J. D}\ }\textbf {\bibinfo {volume} {69}},\ \bibinfo
		{pages} {279} (\bibinfo {year} {2015})},\ \bibinfo {note}
	{arXiv:1508.00442v1}\BibitemShut {NoStop}%
	\bibitem [{\citenamefont {Mahmud}\ and\ \citenamefont
		{Tiesinga}(2013)}]{MahmudTsienga2013}%
	\BibitemOpen
	\bibfield  {author} {\bibinfo {author} {\bibfnamefont {K.~W.}\ \bibnamefont
			{Mahmud}}\ and\ \bibinfo {author} {\bibfnamefont {E.}~\bibnamefont
			{Tiesinga}},\ }\bibfield  {title} {\bibinfo {title} {Dynamics of spin-1
			bosons in an optical lattice: Spin mixing, quantum-phase-revival
			spectroscopy, and effective three-body interactions},\ }\href
	{https://doi.org/10.1103/PhysRevA.88.023602} {\bibfield  {journal} {\bibinfo
			{journal} {Phys. Rev. A}\ }\textbf {\bibinfo {volume} {88}},\ \bibinfo
		{pages} {023602} (\bibinfo {year} {2013})},\ \bibinfo {note}
	{arXiv:1304.7565v1}\BibitemShut {NoStop}%
	\bibitem [{\citenamefont {Meinert}\ \emph {et~al.}(2014)\citenamefont
		{Meinert}, \citenamefont {Mark}, \citenamefont {Kirilov}, \citenamefont
		{Lauber}, \citenamefont {Weinmann}, \citenamefont {Gr\"obner},\ and\
		\citenamefont {N\"agerl}}]{MeinertNaegerl2014}%
	\BibitemOpen
	\bibfield  {author} {\bibinfo {author} {\bibfnamefont {F.}~\bibnamefont
			{Meinert}}, \bibinfo {author} {\bibfnamefont {M.~J.}\ \bibnamefont {Mark}},
		\bibinfo {author} {\bibfnamefont {E.}~\bibnamefont {Kirilov}}, \bibinfo
		{author} {\bibfnamefont {K.}~\bibnamefont {Lauber}}, \bibinfo {author}
		{\bibfnamefont {P.}~\bibnamefont {Weinmann}}, \bibinfo {author}
		{\bibfnamefont {M.}~\bibnamefont {Gr\"obner}},\ and\ \bibinfo {author}
		{\bibfnamefont {H.-C.}\ \bibnamefont {N\"agerl}},\ }\bibfield  {title}
	{\bibinfo {title} {{Interaction-Induced Quantum Phase Revivals and Evidence
				for the Transition to the Quantum Chaotic Regime in 1D Atomic Bloch
				Oscillations}},\ }\href {https://doi.org/10.1103/PhysRevLett.112.193003}
	{\bibfield  {journal} {\bibinfo  {journal} {Phys. Rev. Lett.}\ }\textbf
		{\bibinfo {volume} {112}},\ \bibinfo {pages} {193003} (\bibinfo {year}
		{2014})},\ \bibinfo {note} {arXiv:1309.4045v2}\BibitemShut {NoStop}%
	\bibitem [{\citenamefont {Fan}\ \emph {et~al.}(2019)\citenamefont {Fan},
		\citenamefont {Zhang},\ and\ \citenamefont {Wu}}]{FanWu2019}%
	\BibitemOpen
	\bibfield  {author} {\bibinfo {author} {\bibfnamefont {C.-H.}\ \bibnamefont
			{Fan}}, \bibinfo {author} {\bibfnamefont {H.-X.}\ \bibnamefont {Zhang}},\
		and\ \bibinfo {author} {\bibfnamefont {J.-H.}\ \bibnamefont {Wu}},\
	}\bibfield  {title} {\bibinfo {title} {In-phase and antiphase dynamics of
			rydberg atoms with distinguishable resonances},\ }\href
	{https://doi.org/10.1103/PhysRevA.99.033813} {\bibfield  {journal} {\bibinfo
			{journal} {Phys. Rev. A}\ }\textbf {\bibinfo {volume} {99}},\ \bibinfo
		{pages} {033813} (\bibinfo {year} {2019})}\BibitemShut {NoStop}%
	\bibitem [{\citenamefont {Fan}\ \emph {et~al.}(2020)\citenamefont {Fan},
		\citenamefont {Rossini}, \citenamefont {Zhang}, \citenamefont {Wu},
		\citenamefont {Artoni},\ and\ \citenamefont {La~Rocca}}]{FanRossini2020}%
	\BibitemOpen
	\bibfield  {author} {\bibinfo {author} {\bibfnamefont {C.-H.}\ \bibnamefont
			{Fan}}, \bibinfo {author} {\bibfnamefont {D.}~\bibnamefont {Rossini}},
		\bibinfo {author} {\bibfnamefont {H.-X.}\ \bibnamefont {Zhang}}, \bibinfo
		{author} {\bibfnamefont {J.-H.}\ \bibnamefont {Wu}}, \bibinfo {author}
		{\bibfnamefont {M.}~\bibnamefont {Artoni}},\ and\ \bibinfo {author}
		{\bibfnamefont {G.~C.}\ \bibnamefont {La~Rocca}},\ }\bibfield  {title}
	{\bibinfo {title} {Discrete time crystal in a finite chain of {R}ydberg atoms
			without disorder},\ }\href {https://doi.org/10.1103/PhysRevA.101.013417}
	{\bibfield  {journal} {\bibinfo  {journal} {Phys. Rev. A}\ }\textbf {\bibinfo
			{volume} {101}},\ \bibinfo {pages} {013417} (\bibinfo {year} {2020})},\
	\bibinfo {note} {arXiv:1907.03446v2}\BibitemShut {NoStop}%
	\bibitem [{\citenamefont {Bhaktavatsala~Rao}\ and\ \citenamefont
		{M\o{}lmer}(2014)}]{RaoMoelmer2014}%
	\BibitemOpen
	\bibfield  {author} {\bibinfo {author} {\bibfnamefont {D.~D.}\ \bibnamefont
			{Bhaktavatsala~Rao}}\ and\ \bibinfo {author} {\bibfnamefont {K.}~\bibnamefont
			{M\o{}lmer}},\ }\bibfield  {title} {\bibinfo {title} {{Robust
				{R}ydberg-interaction gates with adiabatic passage}},\ }\href
	{https://doi.org/10.1103/PhysRevA.89.030301} {\bibfield  {journal} {\bibinfo
			{journal} {Phys. Rev. A}\ }\textbf {\bibinfo {volume} {89}},\ \bibinfo
		{pages} {030301} (\bibinfo {year} {2014})},\ \bibinfo {note}
	{arXiv:1311.5147v1}\BibitemShut {NoStop}%
	\bibitem [{\citenamefont {Beterov}\ \emph {et~al.}(2020)\citenamefont
		{Beterov}, \citenamefont {Tretyakov}, \citenamefont {Entin}, \citenamefont
		{Yakshina}, \citenamefont {Ryabtsev}, \citenamefont {Saffman},\ and\
		\citenamefont {Bergamini}}]{BeterovBergamini2020}%
	\BibitemOpen
	\bibfield  {author} {\bibinfo {author} {\bibfnamefont {I.~I.}\ \bibnamefont
			{Beterov}}, \bibinfo {author} {\bibfnamefont {D.~B.}\ \bibnamefont
			{Tretyakov}}, \bibinfo {author} {\bibfnamefont {V.~M.}\ \bibnamefont
			{Entin}}, \bibinfo {author} {\bibfnamefont {E.~A.}\ \bibnamefont {Yakshina}},
		\bibinfo {author} {\bibfnamefont {I.~I.}\ \bibnamefont {Ryabtsev}}, \bibinfo
		{author} {\bibfnamefont {M.}~\bibnamefont {Saffman}},\ and\ \bibinfo {author}
		{\bibfnamefont {S.}~\bibnamefont {Bergamini}},\ }\bibfield  {title} {\bibinfo
		{title} {Application of adiabatic passage in {R}ydberg atomic ensembles for
			quantum information processing},\ }\href
	{https://doi.org/10.1088/1361-6455/ab8719} {\bibfield  {journal} {\bibinfo
			{journal} {J. Phys. B: At. Mol. Opt. Phys.}\ }\textbf {\bibinfo {volume}
			{53}},\ \bibinfo {pages} {182001} (\bibinfo {year} {2020})},\ \bibinfo {note}
	{arXiv:2001.06352v1}\BibitemShut {NoStop}%
	\bibitem [{\citenamefont {de~L\'es\'eleuc}\ \emph {et~al.}(2018)\citenamefont
		{de~L\'es\'eleuc}, \citenamefont {Barredo}, \citenamefont {Lienhard},
		\citenamefont {Browaeys},\ and\ \citenamefont
		{Lahaye}}]{DeLeseleucLahaye2018}%
	\BibitemOpen
	\bibfield  {author} {\bibinfo {author} {\bibfnamefont {S.}~\bibnamefont
			{de~L\'es\'eleuc}}, \bibinfo {author} {\bibfnamefont {D.}~\bibnamefont
			{Barredo}}, \bibinfo {author} {\bibfnamefont {V.}~\bibnamefont {Lienhard}},
		\bibinfo {author} {\bibfnamefont {A.}~\bibnamefont {Browaeys}},\ and\
		\bibinfo {author} {\bibfnamefont {T.}~\bibnamefont {Lahaye}},\ }\bibfield
	{title} {\bibinfo {title} {Analysis of imperfections in the coherent optical
			excitation of single atoms to {R}ydberg states},\ }\href
	{https://doi.org/10.1103/PhysRevA.97.053803} {\bibfield  {journal} {\bibinfo
			{journal} {Phys. Rev. A}\ }\textbf {\bibinfo {volume} {97}},\ \bibinfo
		{pages} {053803} (\bibinfo {year} {2018})},\ \bibinfo {note}
	{arXiv:1802.10424v1}\BibitemShut {NoStop}%
	\bibitem [{\citenamefont {Walls}\ and\ \citenamefont {Milburn}(1995)}]{WM1995}%
	\BibitemOpen
	\bibfield  {author} {\bibinfo {author} {\bibfnamefont {D.~F.}\ \bibnamefont
			{Walls}}\ and\ \bibinfo {author} {\bibfnamefont {G.~J.}\ \bibnamefont
			{Milburn}},\ }\href@noop {} {{\selectlanguage {English}\emph {\bibinfo
				{title} {Quantum optics / D.F. Walls, G.J. Milburn}}}},\ \bibinfo {edition}
	{springer study ed.}\ ed.\ (\bibinfo  {publisher} {Springer-Verlag Berlin ;
		New York},\ \bibinfo {year} {1995})\ pp.\ \bibinfo {pages} {xii, 351 p.
		:}\BibitemShut {NoStop}%
	\bibitem [{\citenamefont {Egan}\ \emph {et~al.}(2021)\citenamefont {Egan},
		\citenamefont {Debroy}, \citenamefont {Noel}, \citenamefont {Risinger},
		\citenamefont {Zhu}, \citenamefont {Biswas}, \citenamefont {Newman},
		\citenamefont {Li}, \citenamefont {Brown}, \citenamefont {Cetina},\ and\
		\citenamefont {Monroe}}]{Monroe2021}%
	\BibitemOpen
	\bibfield  {author} {\bibinfo {author} {\bibfnamefont {L.}~\bibnamefont
			{Egan}}, \bibinfo {author} {\bibfnamefont {D.~M.}\ \bibnamefont {Debroy}},
		\bibinfo {author} {\bibfnamefont {C.}~\bibnamefont {Noel}}, \bibinfo {author}
		{\bibfnamefont {A.}~\bibnamefont {Risinger}}, \bibinfo {author}
		{\bibfnamefont {D.}~\bibnamefont {Zhu}}, \bibinfo {author} {\bibfnamefont
			{D.}~\bibnamefont {Biswas}}, \bibinfo {author} {\bibfnamefont
			{M.}~\bibnamefont {Newman}}, \bibinfo {author} {\bibfnamefont
			{M.}~\bibnamefont {Li}}, \bibinfo {author} {\bibfnamefont {K.~R.}\
			\bibnamefont {Brown}}, \bibinfo {author} {\bibfnamefont {M.}~\bibnamefont
			{Cetina}},\ and\ \bibinfo {author} {\bibfnamefont {C.}~\bibnamefont
			{Monroe}},\ }\bibfield  {title} {\bibinfo {title} {Fault-tolerant control of
			an error-corrected qubit},\ }\href
	{https://doi.org/10.1038/s41586-021-03928-y} {\bibfield  {journal} {\bibinfo
			{journal} {Nature}\ }\textbf {\bibinfo {volume} {598}},\ \bibinfo {pages}
		{281} (\bibinfo {year} {2021})}\BibitemShut {NoStop}%
	\bibitem [{\citenamefont {Saffman}\ \emph
		{et~al.}(2010{\natexlab{a}})\citenamefont {Saffman}, \citenamefont {Walker},\
		and\ \citenamefont {M\o{}lmer}}]{SaffmanMolmer2010}%
	\BibitemOpen
	\bibfield  {author} {\bibinfo {author} {\bibfnamefont {M.}~\bibnamefont
			{Saffman}}, \bibinfo {author} {\bibfnamefont {T.~G.}\ \bibnamefont
			{Walker}},\ and\ \bibinfo {author} {\bibfnamefont {K.}~\bibnamefont
			{M\o{}lmer}},\ }\bibfield  {title} {\bibinfo {title} {Quantum information
			with {R}ydberg atoms},\ }\href {https://doi.org/10.1103/RevModPhys.82.2313}
	{\bibfield  {journal} {\bibinfo  {journal} {Rev. Mod. Phys.}\ }\textbf
		{\bibinfo {volume} {82}},\ \bibinfo {pages} {2313} (\bibinfo {year}
		{2010}{\natexlab{a}})},\ \bibinfo {note} {arXiv:0909.4777v3}\BibitemShut
	{NoStop}%
	\bibitem [{\citenamefont {Labuhn}\ \emph {et~al.}(2016)\citenamefont {Labuhn},
		\citenamefont {Barredo}, \citenamefont {Ravets}, \citenamefont
		{de~L\'es\'eleuc}, \citenamefont {Macr\`i}, \citenamefont {Lahaye},\ and\
		\citenamefont {Browaeys}}]{LabuhnBrowaeys2016}%
	\BibitemOpen
	\bibfield  {author} {\bibinfo {author} {\bibfnamefont {H.}~\bibnamefont
			{Labuhn}}, \bibinfo {author} {\bibfnamefont {D.}~\bibnamefont {Barredo}},
		\bibinfo {author} {\bibfnamefont {S.}~\bibnamefont {Ravets}}, \bibinfo
		{author} {\bibfnamefont {S.}~\bibnamefont {de~L\'es\'eleuc}}, \bibinfo
		{author} {\bibfnamefont {T.}~\bibnamefont {Macr\`i}}, \bibinfo {author}
		{\bibfnamefont {T.}~\bibnamefont {Lahaye}},\ and\ \bibinfo {author}
		{\bibfnamefont {A.}~\bibnamefont {Browaeys}},\ }\bibfield  {title} {\bibinfo
		{title} {Tunable two-dimensional arrays of single {R}ydberg atoms for
			realizing quantum ising models},\ }\bibfield  {journal} {\bibinfo  {journal}
		{Nature}\ }\textbf {\bibinfo {volume} {534}},\ \href
	{https://doi.org/10.1038/nature18274} {10.1038/nature18274} (\bibinfo {year}
	{2016}),\ \bibinfo {note} {arXiv:1509.04543v3}\BibitemShut {NoStop}%
	\bibitem [{\citenamefont {Adams}\ \emph {et~al.}(2019)\citenamefont {Adams},
		\citenamefont {Pritchard},\ and\ \citenamefont {Shaffer}}]{Adams2019}%
	\BibitemOpen
	\bibfield  {author} {\bibinfo {author} {\bibfnamefont {C.~S.}\ \bibnamefont
			{Adams}}, \bibinfo {author} {\bibfnamefont {J.~D.}\ \bibnamefont
			{Pritchard}},\ and\ \bibinfo {author} {\bibfnamefont {J.~P.}\ \bibnamefont
			{Shaffer}},\ }\bibfield  {title} {\bibinfo {title} {{R}ydberg atom quantum
			technologies},\ }\href {https://doi.org/10.1088/1361-6455/ab52ef} {\bibfield
		{journal} {\bibinfo  {journal} {J. Phys. B: At. Mol. Opt. Phys.}\ }\textbf
		{\bibinfo {volume} {53}},\ \bibinfo {pages} {012002} (\bibinfo {year}
		{2019})},\ \bibinfo {note} {arXiv:1907.09231}\BibitemShut {NoStop}%
	\bibitem [{\citenamefont {Zhang}\ \emph {et~al.}(2020)\citenamefont {Zhang},
		\citenamefont {Pokorny}, \citenamefont {Li}, \citenamefont {Higgins},
		\citenamefont {P{\"o}schl}, \citenamefont {Lesanovsky},\ and\ \citenamefont
		{Hennrich}}]{Hennrich2020}%
	\BibitemOpen
	\bibfield  {author} {\bibinfo {author} {\bibfnamefont {C.}~\bibnamefont
			{Zhang}}, \bibinfo {author} {\bibfnamefont {F.}~\bibnamefont {Pokorny}},
		\bibinfo {author} {\bibfnamefont {W.}~\bibnamefont {Li}}, \bibinfo {author}
		{\bibfnamefont {G.}~\bibnamefont {Higgins}}, \bibinfo {author} {\bibfnamefont
			{A.}~\bibnamefont {P{\"o}schl}}, \bibinfo {author} {\bibfnamefont
			{I.}~\bibnamefont {Lesanovsky}},\ and\ \bibinfo {author} {\bibfnamefont
			{M.}~\bibnamefont {Hennrich}},\ }\bibfield  {title} {\bibinfo {title}
		{{Submicrosecond entangling gate between trapped ions via Rydberg
				interaction}},\ }\href {https://doi.org/10.1038/s41586-020-2152-9} {\bibfield
		{journal} {\bibinfo  {journal} {Nature}\ }\textbf {\bibinfo {volume}
			{580}},\ \bibinfo {pages} {345} (\bibinfo {year} {2020})},\ \bibinfo {note}
	{arXiv:1908.11284v1}\BibitemShut {NoStop}%
	\bibitem [{\citenamefont {Borish}\ \emph {et~al.}(2020)\citenamefont {Borish},
		\citenamefont {Markovi\'{c}}, \citenamefont {Hines}, \citenamefont
		{Rajagopal},\ and\ \citenamefont
		{Schleier-Smith}}]{BorishSchleier-Smith2020}%
	\BibitemOpen
	\bibfield  {author} {\bibinfo {author} {\bibfnamefont {V.}~\bibnamefont
			{Borish}}, \bibinfo {author} {\bibfnamefont {O.}~\bibnamefont
			{Markovi\'{c}}}, \bibinfo {author} {\bibfnamefont {J.~A.}\ \bibnamefont
			{Hines}}, \bibinfo {author} {\bibfnamefont {S.~V.}\ \bibnamefont
			{Rajagopal}},\ and\ \bibinfo {author} {\bibfnamefont {M.}~\bibnamefont
			{Schleier-Smith}},\ }\bibfield  {title} {\bibinfo {title} {Transverse-field
			{I}sing dynamics in a {R}ydberg-dressed atomic gas},\ }\href
	{https://doi.org/10.1103/PhysRevLett.124.063601} {\bibfield  {journal}
		{\bibinfo  {journal} {Phys. Rev. Lett.}\ }\textbf {\bibinfo {volume} {124}},\
		\bibinfo {pages} {063601} (\bibinfo {year} {2020})},\ \bibinfo {note}
	{arXiv:1910.13687v3}\BibitemShut {NoStop}%
	\bibitem [{\citenamefont {Ravets}\ \emph {et~al.}(2014)\citenamefont {Ravets},
		\citenamefont {Labuhn}, \citenamefont {Barredo}, \citenamefont {B\'eguin},
		\citenamefont {Lahaye},\ and\ \citenamefont {Browaeys}}]{RavetsBrowaeys2014}%
	\BibitemOpen
	\bibfield  {author} {\bibinfo {author} {\bibfnamefont {S.}~\bibnamefont
			{Ravets}}, \bibinfo {author} {\bibfnamefont {H.}~\bibnamefont {Labuhn}},
		\bibinfo {author} {\bibfnamefont {D.}~\bibnamefont {Barredo}}, \bibinfo
		{author} {\bibfnamefont {L.}~\bibnamefont {B\'eguin}}, \bibinfo {author}
		{\bibfnamefont {T.}~\bibnamefont {Lahaye}},\ and\ \bibinfo {author}
		{\bibfnamefont {A.}~\bibnamefont {Browaeys}},\ }\bibfield  {title} {\bibinfo
		{title} {Coherent dipole-dipole coupling between two single {R}ydberg atoms
			at an electrically-tuned {F}\"orster resonance},\ }\href
	{https://doi.org/10.1038/nphys3119} {\bibfield  {journal} {\bibinfo
			{journal} {Nature Physics}\ }\textbf {\bibinfo {volume} {10}},\ \bibinfo
		{pages} {914} (\bibinfo {year} {2014})},\ \bibinfo {note}
	{arXiv:1405.7804v1}\BibitemShut {NoStop}%
	\bibitem [{\citenamefont {Higgins}\ \emph {et~al.}(2017)\citenamefont
		{Higgins}, \citenamefont {Pokorny}, \citenamefont {Zhang}, \citenamefont
		{Bodart},\ and\ \citenamefont {Hennrich}}]{Hennrich2017}%
	\BibitemOpen
	\bibfield  {author} {\bibinfo {author} {\bibfnamefont {G.}~\bibnamefont
			{Higgins}}, \bibinfo {author} {\bibfnamefont {F.}~\bibnamefont {Pokorny}},
		\bibinfo {author} {\bibfnamefont {C.}~\bibnamefont {Zhang}}, \bibinfo
		{author} {\bibfnamefont {Q.}~\bibnamefont {Bodart}},\ and\ \bibinfo {author}
		{\bibfnamefont {M.}~\bibnamefont {Hennrich}},\ }\bibfield  {title} {\bibinfo
		{title} {Coherent control of a single trapped rydberg ion},\ }\href
	{https://doi.org/10.1103/PhysRevLett.119.220501} {\bibfield  {journal}
		{\bibinfo  {journal} {Phys. Rev. Lett.}\ }\textbf {\bibinfo {volume} {119}},\
		\bibinfo {pages} {220501} (\bibinfo {year} {2017})},\ \bibinfo {note}
	{arXiv:1708.06387}\BibitemShut {NoStop}%
	\bibitem [{\citenamefont {Huang}\ \emph {et~al.}(2018)\citenamefont {Huang},
		\citenamefont {Ding}, \citenamefont {Hu}, \citenamefont {Shen}, \citenamefont
		{Li}, \citenamefont {Wu},\ and\ \citenamefont {Zheng}}]{HuangZheng2018}%
	\BibitemOpen
	\bibfield  {author} {\bibinfo {author} {\bibfnamefont {X.-R.}\ \bibnamefont
			{Huang}}, \bibinfo {author} {\bibfnamefont {Z.-X.}\ \bibnamefont {Ding}},
		\bibinfo {author} {\bibfnamefont {C.-S.}\ \bibnamefont {Hu}}, \bibinfo
		{author} {\bibfnamefont {L.-T.}\ \bibnamefont {Shen}}, \bibinfo {author}
		{\bibfnamefont {W.}~\bibnamefont {Li}}, \bibinfo {author} {\bibfnamefont
			{H.}~\bibnamefont {Wu}},\ and\ \bibinfo {author} {\bibfnamefont {S.-B.}\
			\bibnamefont {Zheng}},\ }\bibfield  {title} {\bibinfo {title} {Robust
			{R}ydberg gate via {L}andau-{Z}ener control of {F}\"orster resonance},\
	}\href {https://doi.org/10.1103/PhysRevA.98.052324} {\bibfield  {journal}
		{\bibinfo  {journal} {Phys. Rev. A}\ }\textbf {\bibinfo {volume} {98}},\
		\bibinfo {pages} {052324} (\bibinfo {year} {2018})},\ \bibinfo {note}
	{arXiv:1806.09775v4}\BibitemShut {NoStop}%
	\bibitem [{\citenamefont {Schwartz}\ \emph {et~al.}(2021)\citenamefont
		{Schwartz}, \citenamefont {Shimazaki}, \citenamefont {Kuhlenkamp},
		\citenamefont {Watanabe}, \citenamefont {Taniguchi}, \citenamefont {Kroner},\
		and\ \citenamefont {Imamoglu}}]{Atac2021}%
	\BibitemOpen
	\bibfield  {author} {\bibinfo {author} {\bibfnamefont {I.}~\bibnamefont
			{Schwartz}}, \bibinfo {author} {\bibfnamefont {Y.}~\bibnamefont {Shimazaki}},
		\bibinfo {author} {\bibfnamefont {C.}~\bibnamefont {Kuhlenkamp}}, \bibinfo
		{author} {\bibfnamefont {K.}~\bibnamefont {Watanabe}}, \bibinfo {author}
		{\bibfnamefont {T.}~\bibnamefont {Taniguchi}}, \bibinfo {author}
		{\bibfnamefont {M.}~\bibnamefont {Kroner}},\ and\ \bibinfo {author}
		{\bibfnamefont {A.}~\bibnamefont {Imamoglu}},\ }\bibfield  {title} {\bibinfo
		{title} {Electrically tunable feshbach resonances in twisted bilayer
			semiconductors},\ }\href {https://doi.org/10.1126/science.abj3831} {\bibfield
		{journal} {\bibinfo  {journal} {Science}\ }\textbf {\bibinfo {volume}
			{374}},\ \bibinfo {pages} {336} (\bibinfo {year} {2021})},\ \Eprint
	{https://arxiv.org/abs/https://www.science.org/doi/pdf/10.1126/science.abj3831}
	{https://www.science.org/doi/pdf/10.1126/science.abj3831} \BibitemShut
	{NoStop}%
	\bibitem [{\citenamefont {Taherkhani}\ \emph {et~al.}(2019)\citenamefont
		{Taherkhani}, \citenamefont {Willatzen}, \citenamefont {Denning},
		\citenamefont {Protsenko},\ and\ \citenamefont
		{Gregersen}}]{TaherkhaniGregersen2019}%
	\BibitemOpen
	\bibfield  {author} {\bibinfo {author} {\bibfnamefont {M.}~\bibnamefont
			{Taherkhani}}, \bibinfo {author} {\bibfnamefont {M.}~\bibnamefont
			{Willatzen}}, \bibinfo {author} {\bibfnamefont {E.~V.}\ \bibnamefont
			{Denning}}, \bibinfo {author} {\bibfnamefont {I.~E.}\ \bibnamefont
			{Protsenko}},\ and\ \bibinfo {author} {\bibfnamefont {N.}~\bibnamefont
			{Gregersen}},\ }\bibfield  {title} {\bibinfo {title} {High-fidelity optical
			quantum gates based on type-ii double quantum dots in a nanowire},\ }\href
	{https://doi.org/10.1103/PhysRevB.99.165305} {\bibfield  {journal} {\bibinfo
			{journal} {Phys. Rev. B}\ }\textbf {\bibinfo {volume} {99}},\ \bibinfo
		{pages} {165305} (\bibinfo {year} {2019})}\BibitemShut {NoStop}%
	\bibitem [{\citenamefont {Giannelli}\ and\ \citenamefont
		{Arimondo}(2014)}]{GiannelliArimondo2014}%
	\BibitemOpen
	\bibfield  {author} {\bibinfo {author} {\bibfnamefont {L.}~\bibnamefont
			{Giannelli}}\ and\ \bibinfo {author} {\bibfnamefont {E.}~\bibnamefont
			{Arimondo}},\ }\bibfield  {title} {\bibinfo {title} {Three-level
			superadiabatic quantum driving},\ }\href
	{https://doi.org/10.1103/PhysRevA.89.033419} {\bibfield  {journal} {\bibinfo
			{journal} {Phys. Rev. A}\ }\textbf {\bibinfo {volume} {89}},\ \bibinfo
		{pages} {033419} (\bibinfo {year} {2014})},\ \bibinfo {note}
	{arXiv:1402.1299v1}\BibitemShut {NoStop}%
	\bibitem [{\citenamefont {Vitanov}\ \emph {et~al.}(2017)\citenamefont
		{Vitanov}, \citenamefont {Rangelov}, \citenamefont {Shore},\ and\
		\citenamefont {Bergmann}}]{VitanovBergmann2017}%
	\BibitemOpen
	\bibfield  {author} {\bibinfo {author} {\bibfnamefont {N.~V.}\ \bibnamefont
			{Vitanov}}, \bibinfo {author} {\bibfnamefont {A.~A.}\ \bibnamefont
			{Rangelov}}, \bibinfo {author} {\bibfnamefont {B.~W.}\ \bibnamefont
			{Shore}},\ and\ \bibinfo {author} {\bibfnamefont {K.}~\bibnamefont
			{Bergmann}},\ }\bibfield  {title} {\bibinfo {title} {Stimulated {R}aman
			adiabatic passage in physics, chemistry, and beyond},\ }\href
	{https://doi.org/10.1103/RevModPhys.89.015006} {\bibfield  {journal}
		{\bibinfo  {journal} {Rev. Mod. Phys.}\ }\textbf {\bibinfo {volume} {89}},\
		\bibinfo {pages} {015006} (\bibinfo {year} {2017})},\ \bibinfo {note}
	{arXiv:1605.00224v2}\BibitemShut {NoStop}%
	\bibitem [{\citenamefont {Beterov}\ \emph {et~al.}(2013)\citenamefont
		{Beterov}, \citenamefont {Saffman}, \citenamefont {Yakshina}, \citenamefont
		{Zhukov}, \citenamefont {Tretyakov}, \citenamefont {Entin}, \citenamefont
		{Ryabtsev}, \citenamefont {Mansell}, \citenamefont {MacCormick},
		\citenamefont {Bergamini},\ and\ \citenamefont
		{Fedoruk}}]{BeterovSaffman2013}%
	\BibitemOpen
	\bibfield  {author} {\bibinfo {author} {\bibfnamefont {I.~I.}\ \bibnamefont
			{Beterov}}, \bibinfo {author} {\bibfnamefont {M.}~\bibnamefont {Saffman}},
		\bibinfo {author} {\bibfnamefont {E.~A.}\ \bibnamefont {Yakshina}}, \bibinfo
		{author} {\bibfnamefont {V.~P.}\ \bibnamefont {Zhukov}}, \bibinfo {author}
		{\bibfnamefont {D.~B.}\ \bibnamefont {Tretyakov}}, \bibinfo {author}
		{\bibfnamefont {V.~M.}\ \bibnamefont {Entin}}, \bibinfo {author}
		{\bibfnamefont {I.~I.}\ \bibnamefont {Ryabtsev}}, \bibinfo {author}
		{\bibfnamefont {C.~W.}\ \bibnamefont {Mansell}}, \bibinfo {author}
		{\bibfnamefont {C.}~\bibnamefont {MacCormick}}, \bibinfo {author}
		{\bibfnamefont {S.}~\bibnamefont {Bergamini}},\ and\ \bibinfo {author}
		{\bibfnamefont {M.~P.}\ \bibnamefont {Fedoruk}},\ }\bibfield  {title}
	{\bibinfo {title} {Quantum gates in mesoscopic atomic ensembles based on
			adiabatic passage and {R}ydberg blockade},\ }\href
	{https://doi.org/10.1103/PhysRevA.88.010303} {\bibfield  {journal} {\bibinfo
			{journal} {Phys. Rev. A}\ }\textbf {\bibinfo {volume} {88}},\ \bibinfo
		{pages} {010303} (\bibinfo {year} {2013})},\ \bibinfo {note}
	{arXiv:1212.1138v3}\BibitemShut {NoStop}%
	\bibitem [{\citenamefont {Laine}\ and\ \citenamefont
		{Stenholm}(1996)}]{LaineStenholm1996}%
	\BibitemOpen
	\bibfield  {author} {\bibinfo {author} {\bibfnamefont {T.~A.}\ \bibnamefont
			{Laine}}\ and\ \bibinfo {author} {\bibfnamefont {S.}~\bibnamefont
			{Stenholm}},\ }\bibfield  {title} {\bibinfo {title} {Adiabatic processes in
			three-level systems},\ }\href {https://doi.org/10.1103/PhysRevA.53.2501}
	{\bibfield  {journal} {\bibinfo  {journal} {Phys. Rev. A}\ }\textbf {\bibinfo
			{volume} {53}},\ \bibinfo {pages} {2501} (\bibinfo {year}
		{1996})}\BibitemShut {NoStop}%
	\bibitem [{\citenamefont {Vitanov}\ and\ \citenamefont
		{Stenholm}(1996)}]{VitanovStenholm1996}%
	\BibitemOpen
	\bibfield  {author} {\bibinfo {author} {\bibfnamefont {N.~V.}\ \bibnamefont
			{Vitanov}}\ and\ \bibinfo {author} {\bibfnamefont {S.}~\bibnamefont
			{Stenholm}},\ }\bibfield  {title} {\bibinfo {title} {Non-adiabatic effects in
			population transfer in three-level systems},\ }\href
	{https://doi.org/https://doi.org/10.1016/0030-4018(96)00216-7} {\bibfield
		{journal} {\bibinfo  {journal} {Opt. Comm.}\ }\textbf {\bibinfo {volume}
			{127}},\ \bibinfo {pages} {215 } (\bibinfo {year} {1996})}\BibitemShut
	{NoStop}%
	\bibitem [{\citenamefont {Torosov}\ and\ \citenamefont
		{Vitanov}(2013)}]{TorosovVitanov2013}%
	\BibitemOpen
	\bibfield  {author} {\bibinfo {author} {\bibfnamefont {B.~T.}\ \bibnamefont
			{Torosov}}\ and\ \bibinfo {author} {\bibfnamefont {N.~V.}\ \bibnamefont
			{Vitanov}},\ }\bibfield  {title} {\bibinfo {title} {Composite stimulated
			{R}aman adiabatic passage},\ }\href
	{https://doi.org/10.1103/PhysRevA.87.043418} {\bibfield  {journal} {\bibinfo
			{journal} {Phys. Rev. A}\ }\textbf {\bibinfo {volume} {87}},\ \bibinfo
		{pages} {043418} (\bibinfo {year} {2013})},\ \bibinfo {note}
	{arXiv:1306.0699v1}\BibitemShut {NoStop}%
	\bibitem [{\citenamefont {Shore}(2011)}]{ShoreBook}%
	\BibitemOpen
	\bibfield  {author} {\bibinfo {author} {\bibfnamefont {B.}~\bibnamefont
			{Shore}},\ }\href@noop {} {{\selectlanguage {English}\emph {\bibinfo {title}
				{Manipulating quantum structures using laser pulses}}}}\ (\bibinfo
	{publisher} {Cambridge University Press},\ \bibinfo {year}
	{2011})\BibitemShut {NoStop}%
	\bibitem [{\citenamefont {Ficek}\ and\ \citenamefont
		{Tana\'{s}}(2002)}]{FicekTanas2002}%
	\BibitemOpen
	\bibfield  {author} {\bibinfo {author} {\bibfnamefont {Z.}~\bibnamefont
			{Ficek}}\ and\ \bibinfo {author} {\bibfnamefont {R.}~\bibnamefont
			{Tana\'{s}}},\ }\bibfield  {title} {\bibinfo {title} {Entangled states and
			collective nonclassical effects in two-atom systems},\ }\href
	{https://doi.org/https://doi.org/10.1016/S0370-1573(02)00368-X} {\bibfield
		{journal} {\bibinfo  {journal} {Phys. Rep.}\ }\textbf {\bibinfo {volume}
			{372}},\ \bibinfo {pages} {369 } (\bibinfo {year} {2002})},\ \bibinfo {note}
	{arXiv:quant-ph/0302082v1}\BibitemShut {NoStop}%
	\bibitem [{\citenamefont {Comparat}\ and\ \citenamefont
		{Pillet}(2010)}]{ComparatPillet2010}%
	\BibitemOpen
	\bibfield  {author} {\bibinfo {author} {\bibfnamefont {D.}~\bibnamefont
			{Comparat}}\ and\ \bibinfo {author} {\bibfnamefont {P.}~\bibnamefont
			{Pillet}},\ }\bibfield  {title} {\bibinfo {title} {Dipole blockade in a cold
			{R}ydberg atomic sample},\ }\href {https://doi.org/10.1364/JOSAB.27.00A208}
	{\bibfield  {journal} {\bibinfo  {journal} {J. Opt. Soc. Am. B}\ }\textbf
		{\bibinfo {volume} {27}},\ \bibinfo {pages} {A208} (\bibinfo {year}
		{2010})},\ \bibinfo {note} {arXiv:1006.0742v1}\BibitemShut {NoStop}%
	\bibitem [{\citenamefont {Almutairi}\ \emph {et~al.}(2011)\citenamefont
		{Almutairi}, \citenamefont {Tana\ifmmode~\acute{s}\else \'{s}\fi{}},\ and\
		\citenamefont {Ficek}}]{AlmutairiFicek2011}%
	\BibitemOpen
	\bibfield  {author} {\bibinfo {author} {\bibfnamefont {K.}~\bibnamefont
			{Almutairi}}, \bibinfo {author} {\bibfnamefont {R.}~\bibnamefont
			{Tana\ifmmode~\acute{s}\else \'{s}\fi{}}},\ and\ \bibinfo {author}
		{\bibfnamefont {Z.}~\bibnamefont {Ficek}},\ }\bibfield  {title} {\bibinfo
		{title} {Generating two-photon entangled states in a driven two-atom
			system},\ }\href {https://doi.org/10.1103/PhysRevA.84.013831} {\bibfield
		{journal} {\bibinfo  {journal} {Phys. Rev. A}\ }\textbf {\bibinfo {volume}
			{84}},\ \bibinfo {pages} {013831} (\bibinfo {year} {2011})}\BibitemShut
	{NoStop}%
	\bibitem [{\citenamefont {Ates}\ \emph {et~al.}(2007)\citenamefont {Ates},
		\citenamefont {Pohl}, \citenamefont {Pattard},\ and\ \citenamefont
		{Rost}}]{AtosRost2007}%
	\BibitemOpen
	\bibfield  {author} {\bibinfo {author} {\bibfnamefont {C.}~\bibnamefont
			{Ates}}, \bibinfo {author} {\bibfnamefont {T.}~\bibnamefont {Pohl}}, \bibinfo
		{author} {\bibfnamefont {T.}~\bibnamefont {Pattard}},\ and\ \bibinfo {author}
		{\bibfnamefont {J.~M.}\ \bibnamefont {Rost}},\ }\bibfield  {title} {\bibinfo
		{title} {Antiblockade in {R}ydberg excitation of an ultracold lattice gas},\
	}\href {https://doi.org/10.1103/PhysRevLett.98.023002} {\bibfield  {journal}
		{\bibinfo  {journal} {Phys. Rev. Lett.}\ }\textbf {\bibinfo {volume} {98}},\
		\bibinfo {pages} {023002} (\bibinfo {year} {2007})},\ \bibinfo {note}
	{arXiv:physics/0605111v1}\BibitemShut {NoStop}%
	\bibitem [{\citenamefont {Amthor}\ \emph {et~al.}(2010)\citenamefont {Amthor},
		\citenamefont {Giese}, \citenamefont {Hofmann},\ and\ \citenamefont
		{Weidem\"uller}}]{AmthorWeidemuller2010}%
	\BibitemOpen
	\bibfield  {author} {\bibinfo {author} {\bibfnamefont {T.}~\bibnamefont
			{Amthor}}, \bibinfo {author} {\bibfnamefont {C.}~\bibnamefont {Giese}},
		\bibinfo {author} {\bibfnamefont {C.~S.}\ \bibnamefont {Hofmann}},\ and\
		\bibinfo {author} {\bibfnamefont {M.}~\bibnamefont {Weidem\"uller}},\
	}\bibfield  {title} {\bibinfo {title} {Evidence of antiblockade in an
			ultracold {R}ydberg gas},\ }\href
	{https://doi.org/10.1103/PhysRevLett.104.013001} {\bibfield  {journal}
		{\bibinfo  {journal} {Phys. Rev. Lett.}\ }\textbf {\bibinfo {volume} {104}},\
		\bibinfo {pages} {013001} (\bibinfo {year} {2010})},\ \bibinfo {note}
	{arXiv:0909.0837}\BibitemShut {NoStop}%
	\bibitem [{\citenamefont {Kara}\ \emph {et~al.}(2018)\citenamefont {Kara},
		\citenamefont {Bhowmick},\ and\ \citenamefont
		{Mohapatra}}]{KaraMohapatra2018}%
	\BibitemOpen
	\bibfield  {author} {\bibinfo {author} {\bibfnamefont {D.}~\bibnamefont
			{Kara}}, \bibinfo {author} {\bibfnamefont {A.}~\bibnamefont {Bhowmick}},\
		and\ \bibinfo {author} {\bibfnamefont {A.~K.}\ \bibnamefont {Mohapatra}},\
	}\bibfield  {title} {\bibinfo {title} {Rydberg interaction induced enhanced
			excitation in thermal atomic vapor},\ }\href
	{https://doi.org/https://doi.org/10.1038/s41598-018-23559-0} {\bibfield
		{journal} {\bibinfo  {journal} {Scientific Reports}\ }\textbf {\bibinfo
			{volume} {8}},\ \bibinfo {pages} {3236} (\bibinfo {year} {2018})},\ \bibinfo
	{note} {arXiv:1710.05573v1}\BibitemShut {NoStop}%
	\bibitem [{\citenamefont {Arimondo}(1996)}]{Arimondo1996}%
	\BibitemOpen
	\bibfield  {author} {\bibinfo {author} {\bibfnamefont {E.}~\bibnamefont
			{Arimondo}},\ }\bibfield  {title} {\bibinfo {title} {Coherent population
			trapping in laser spectroscopy}\ }(\bibinfo  {publisher} {Elsevier},\
	\bibinfo {year} {1996})\ pp.\ \bibinfo {pages} {257--354}\BibitemShut
	{NoStop}%
	\bibitem [{\citenamefont {Fleischhauer}\ \emph {et~al.}(2005)\citenamefont
		{Fleischhauer}, \citenamefont {Imamoglu},\ and\ \citenamefont
		{Marangos}}]{FleischhauerMarangos2005}%
	\BibitemOpen
	\bibfield  {author} {\bibinfo {author} {\bibfnamefont {M.}~\bibnamefont
			{Fleischhauer}}, \bibinfo {author} {\bibfnamefont {A.}~\bibnamefont
			{Imamoglu}},\ and\ \bibinfo {author} {\bibfnamefont {J.~P.}\ \bibnamefont
			{Marangos}},\ }\bibfield  {title} {\bibinfo {title} {Electromagnetically
			induced transparency: Optics in coherent media},\ }\href
	{https://doi.org/10.1103/RevModPhys.77.633} {\bibfield  {journal} {\bibinfo
			{journal} {Rev. Mod. Phys.}\ }\textbf {\bibinfo {volume} {77}},\ \bibinfo
		{pages} {633} (\bibinfo {year} {2005})}\BibitemShut {NoStop}%
	\bibitem [{\citenamefont {Shore}(2017)}]{Shore2017}%
	\BibitemOpen
	\bibfield  {author} {\bibinfo {author} {\bibfnamefont {B.~W.}\ \bibnamefont
			{Shore}},\ }\bibfield  {title} {\bibinfo {title} {{Picturing stimulated
				{R}aman adiabatic passage: a STIRAP tutorial}},\ }\href
	{https://doi.org/10.1364/AOP.9.000563} {\bibfield  {journal} {\bibinfo
			{journal} {{Adv. Opt. Phot.}}\ }\textbf {\bibinfo {volume} {9}},\ \bibinfo
		{pages} {563} (\bibinfo {year} {2017})}\BibitemShut {NoStop}%
	\bibitem [{\citenamefont {Vitanov}\ and\ \citenamefont
		{Stenholm}(1997)}]{VitanovStenholm1997}%
	\BibitemOpen
	\bibfield  {author} {\bibinfo {author} {\bibfnamefont {N.~V.}\ \bibnamefont
			{Vitanov}}\ and\ \bibinfo {author} {\bibfnamefont {S.}~\bibnamefont
			{Stenholm}},\ }\bibfield  {title} {\bibinfo {title} {Properties of stimulated
			{R}aman adiabatic passage with intermediate-level detuning},\ }\href
	{https://doi.org/https://doi.org/10.1016/S0030-4018(96)00635-9} {\bibfield
		{journal} {\bibinfo  {journal} {Opt. Comm.}\ }\textbf {\bibinfo {volume}
			{135}},\ \bibinfo {pages} {394 } (\bibinfo {year} {1997})}\BibitemShut
	{NoStop}%
	\bibitem [{\citenamefont {Saffman}\ \emph
		{et~al.}(2010{\natexlab{b}})\citenamefont {Saffman}, \citenamefont {Walker},\
		and\ \citenamefont {M\o{}lmer}}]{RevModPhys.82.2313}%
	\BibitemOpen
	\bibfield  {author} {\bibinfo {author} {\bibfnamefont {M.}~\bibnamefont
			{Saffman}}, \bibinfo {author} {\bibfnamefont {T.~G.}\ \bibnamefont
			{Walker}},\ and\ \bibinfo {author} {\bibfnamefont {K.}~\bibnamefont
			{M\o{}lmer}},\ }\bibfield  {title} {\bibinfo {title} {Quantum information
			with rydberg atoms},\ }\href {https://doi.org/10.1103/RevModPhys.82.2313}
	{\bibfield  {journal} {\bibinfo  {journal} {Rev. Mod. Phys.}\ }\textbf
		{\bibinfo {volume} {82}},\ \bibinfo {pages} {2313} (\bibinfo {year}
		{2010}{\natexlab{b}})}\BibitemShut {NoStop}%
	\bibitem [{\citenamefont {Viola}\ \emph {et~al.}(1999)\citenamefont {Viola},
		\citenamefont {Knill},\ and\ \citenamefont {Lloyd}}]{PhysRevLett.82.2417}%
	\BibitemOpen
	\bibfield  {author} {\bibinfo {author} {\bibfnamefont {L.}~\bibnamefont
			{Viola}}, \bibinfo {author} {\bibfnamefont {E.}~\bibnamefont {Knill}},\ and\
		\bibinfo {author} {\bibfnamefont {S.}~\bibnamefont {Lloyd}},\ }\bibfield
	{title} {\bibinfo {title} {Dynamical decoupling of open quantum systems},\
	}\href {https://doi.org/10.1103/PhysRevLett.82.2417} {\bibfield  {journal}
		{\bibinfo  {journal} {Phys. Rev. Lett.}\ }\textbf {\bibinfo {volume} {82}},\
		\bibinfo {pages} {2417} (\bibinfo {year} {1999})},\ \bibinfo {note}
	{arXiv:quant-ph/9809071v2}\BibitemShut {NoStop}%
	\bibitem [{\citenamefont {Lidar}\ and\ \citenamefont
		{Brun}(2013)}]{lidar_brun_2013}%
	\BibitemOpen
	\bibfield  {author} {\bibinfo {author} {\bibfnamefont {D.~A.}\ \bibnamefont
			{Lidar}}\ and\ \bibinfo {author} {\bibfnamefont {T.~A.}\ \bibnamefont
			{Brun}},\ }\href {https://doi.org/10.1017/CBO9781139034807} {\emph {\bibinfo
			{title} {Quantum Error Correction}}}\ (\bibinfo  {publisher} {Cambridge
		University Press},\ \bibinfo {year} {2013})\BibitemShut {NoStop}%
	\bibitem [{\citenamefont {Izrailev}(1990)}]{Izrailev1990}%
	\BibitemOpen
	\bibfield  {author} {\bibinfo {author} {\bibfnamefont {F.~M.}\ \bibnamefont
			{Izrailev}},\ }\bibfield  {title} {\bibinfo {title} {Simple models of quantum
			chaos: Spectrum and eigenfunctions},\ }\href
	{https://doi.org/https://doi.org/10.1016/0370-1573(90)90067-C} {\bibfield
		{journal} {\bibinfo  {journal} {Phys. Rep.}\ }\textbf {\bibinfo {volume}
			{196}},\ \bibinfo {pages} {299} (\bibinfo {year} {1990})}\BibitemShut
	{NoStop}%
	\bibitem [{\citenamefont {Sadgrove}\ and\ \citenamefont
		{Wimberger}(2011)}]{Sadgrove2011}%
	\BibitemOpen
	\bibfield  {author} {\bibinfo {author} {\bibfnamefont {M.}~\bibnamefont
			{Sadgrove}}\ and\ \bibinfo {author} {\bibfnamefont {S.}~\bibnamefont
			{Wimberger}},\ }\bibfield  {title} {\bibinfo {title} {Chapter 7 - {A
				Pseudoclassical Method for the Atom-Optics Kicked Rotor: from Theory to
				Experiment and Back}}\ }(\bibinfo  {publisher} {Academic Press},\ \bibinfo
	{year} {2011})\ pp.\ \bibinfo {pages} {315--369}\BibitemShut {NoStop}%
	\bibitem [{\citenamefont {Jaksch}\ \emph {et~al.}(2000)\citenamefont {Jaksch},
		\citenamefont {Cirac}, \citenamefont {Zoller}, \citenamefont {Rolston},
		\citenamefont {C\^ot\'e},\ and\ \citenamefont {Lukin}}]{PhysRevLett.85.2208}%
	\BibitemOpen
	\bibfield  {author} {\bibinfo {author} {\bibfnamefont {D.}~\bibnamefont
			{Jaksch}}, \bibinfo {author} {\bibfnamefont {J.~I.}\ \bibnamefont {Cirac}},
		\bibinfo {author} {\bibfnamefont {P.}~\bibnamefont {Zoller}}, \bibinfo
		{author} {\bibfnamefont {S.~L.}\ \bibnamefont {Rolston}}, \bibinfo {author}
		{\bibfnamefont {R.}~\bibnamefont {C\^ot\'e}},\ and\ \bibinfo {author}
		{\bibfnamefont {M.~D.}\ \bibnamefont {Lukin}},\ }\bibfield  {title} {\bibinfo
		{title} {Fast quantum gates for neutral atoms},\ }\href
	{https://doi.org/10.1103/PhysRevLett.85.2208} {\bibfield  {journal} {\bibinfo
			{journal} {Phys. Rev. Lett.}\ }\textbf {\bibinfo {volume} {85}},\ \bibinfo
		{pages} {2208} (\bibinfo {year} {2000})}\BibitemShut {NoStop}%
	\bibitem [{\citenamefont {Fedoseev}\ \emph {et~al.}(2021)\citenamefont
		{Fedoseev}, \citenamefont {Luna}, \citenamefont {Hedgepeth}, \citenamefont
		{L\"offler},\ and\ \citenamefont {Bouwmeester}}]{Fedoseev2021}%
	\BibitemOpen
	\bibfield  {author} {\bibinfo {author} {\bibfnamefont {V.}~\bibnamefont
			{Fedoseev}}, \bibinfo {author} {\bibfnamefont {F.}~\bibnamefont {Luna}},
		\bibinfo {author} {\bibfnamefont {I.}~\bibnamefont {Hedgepeth}}, \bibinfo
		{author} {\bibfnamefont {W.}~\bibnamefont {L\"offler}},\ and\ \bibinfo
		{author} {\bibfnamefont {D.}~\bibnamefont {Bouwmeester}},\ }\bibfield
	{title} {\bibinfo {title} {{Stimulated Raman Adiabatic Passage in
				Optomechanics}},\ }\href {https://doi.org/10.1103/PhysRevLett.126.113601}
	{\bibfield  {journal} {\bibinfo  {journal} {Phys. Rev. Lett.}\ }\textbf
		{\bibinfo {volume} {126}},\ \bibinfo {pages} {113601} (\bibinfo {year}
		{2021})},\ \bibinfo {note} {arXiv:1911.11464v2}\BibitemShut {NoStop}%
	\bibitem [{\citenamefont {Saffman}\ \emph {et~al.}(2020)\citenamefont
		{Saffman}, \citenamefont {Beterov}, \citenamefont {Dalal}, \citenamefont
		{P\'aez},\ and\ \citenamefont {Sanders}}]{SaffmanSanders2020}%
	\BibitemOpen
	\bibfield  {author} {\bibinfo {author} {\bibfnamefont {M.}~\bibnamefont
			{Saffman}}, \bibinfo {author} {\bibfnamefont {I.~I.}\ \bibnamefont
			{Beterov}}, \bibinfo {author} {\bibfnamefont {A.}~\bibnamefont {Dalal}},
		\bibinfo {author} {\bibfnamefont {E.~J.}\ \bibnamefont {P\'aez}},\ and\
		\bibinfo {author} {\bibfnamefont {B.~C.}\ \bibnamefont {Sanders}},\
	}\bibfield  {title} {\bibinfo {title} {Symmetric {R}ydberg controlled-$z$
			gates with adiabatic pulses},\ }\href
	{https://doi.org/10.1103/PhysRevA.101.062309} {\bibfield  {journal} {\bibinfo
			{journal} {Phys. Rev. A}\ }\textbf {\bibinfo {volume} {101}},\ \bibinfo
		{pages} {062309} (\bibinfo {year} {2020})},\ \bibinfo {note}
	{arXiv:1912.02977v3}\BibitemShut {NoStop}%
	\bibitem [{\citenamefont {Wu}\ \emph {et~al.}(2021)\citenamefont {Wu},
		\citenamefont {Wang}, \citenamefont {Han}, \citenamefont {Su}, \citenamefont
		{Xia}, \citenamefont {Jiang},\ and\ \citenamefont {Song}}]{Wu2021}%
	\BibitemOpen
	\bibfield  {author} {\bibinfo {author} {\bibfnamefont {J.-L.}\ \bibnamefont
			{Wu}}, \bibinfo {author} {\bibfnamefont {Y.}~\bibnamefont {Wang}}, \bibinfo
		{author} {\bibfnamefont {J.-X.}\ \bibnamefont {Han}}, \bibinfo {author}
		{\bibfnamefont {S.-L.}\ \bibnamefont {Su}}, \bibinfo {author} {\bibfnamefont
			{Y.}~\bibnamefont {Xia}}, \bibinfo {author} {\bibfnamefont {Y.}~\bibnamefont
			{Jiang}},\ and\ \bibinfo {author} {\bibfnamefont {J.}~\bibnamefont {Song}},\
	}\bibfield  {title} {\bibinfo {title} {{Resilient quantum gates on
				periodically driven Rydberg atoms}},\ }\href
	{https://doi.org/10.1103/PhysRevA.103.012601} {\bibfield  {journal} {\bibinfo
			{journal} {Phys. Rev. A}\ }\textbf {\bibinfo {volume} {103}},\ \bibinfo
		{pages} {012601} (\bibinfo {year} {2021})},\ \bibinfo {note}
	{arXiv:2101.02328v1}\BibitemShut {NoStop}%
	\bibitem [{\citenamefont {Di~Stefano}\ \emph {et~al.}(2016)\citenamefont
		{Di~Stefano}, \citenamefont {Paladino}, \citenamefont {Pope},\ and\
		\citenamefont {Falci}}]{DiStefanoFalci2016}%
	\BibitemOpen
	\bibfield  {author} {\bibinfo {author} {\bibfnamefont {P.~G.}\ \bibnamefont
			{Di~Stefano}}, \bibinfo {author} {\bibfnamefont {E.}~\bibnamefont
			{Paladino}}, \bibinfo {author} {\bibfnamefont {T.~J.}\ \bibnamefont {Pope}},\
		and\ \bibinfo {author} {\bibfnamefont {G.}~\bibnamefont {Falci}},\ }\bibfield
	{title} {\bibinfo {title} {Coherent manipulation of noise-protected
			superconducting artificial atoms in the lambda scheme},\ }\href
	{https://doi.org/10.1103/PhysRevA.93.051801} {\bibfield  {journal} {\bibinfo
			{journal} {Phys. Rev. A}\ }\textbf {\bibinfo {volume} {93}},\ \bibinfo
		{pages} {051801} (\bibinfo {year} {2016})},\ \bibinfo {note}
	{arXiv:1509.05562v2}\BibitemShut {NoStop}%
	\bibitem [{\citenamefont {Blais}\ \emph {et~al.}(2021)\citenamefont {Blais},
		\citenamefont {Grimsmo}, \citenamefont {Girvin},\ and\ \citenamefont
		{Wallraff}}]{RevModPhys.93.025005}%
	\BibitemOpen
	\bibfield  {author} {\bibinfo {author} {\bibfnamefont {A.}~\bibnamefont
			{Blais}}, \bibinfo {author} {\bibfnamefont {A.~L.}\ \bibnamefont {Grimsmo}},
		\bibinfo {author} {\bibfnamefont {S.~M.}\ \bibnamefont {Girvin}},\ and\
		\bibinfo {author} {\bibfnamefont {A.}~\bibnamefont {Wallraff}},\ }\bibfield
	{title} {\bibinfo {title} {Circuit quantum electrodynamics},\ }\href
	{https://doi.org/10.1103/RevModPhys.93.025005} {\bibfield  {journal}
		{\bibinfo  {journal} {Rev. Mod. Phys.}\ }\textbf {\bibinfo {volume} {93}},\
		\bibinfo {pages} {025005} (\bibinfo {year} {2021})}\BibitemShut {NoStop}%
	\bibitem [{\citenamefont {Delvecchio}(2021)}]{MathNotebook}%
	\BibitemOpen
	\bibfield  {author} {\bibinfo {author} {\bibfnamefont {M.}~\bibnamefont
			{Delvecchio}},\ }\href {https://notebookarchive.org/2021-06-bjbq4hz}
	{\bibinfo {title} {Two periodically driven interacting qubits}} (\bibinfo
	{year} {2021}),\ \bibinfo {note} {mathematica notebook, available online:
		https://notebookarchive.org/2021-06-bjbq4hz}\BibitemShut {NoStop}%
\end{thebibliography}

%

\end{document}